\documentclass[aps,prd,preprint]{revtex4}

\usepackage{graphicx}
\usepackage{psfrag}

\begin{document}
\title{The Instantaneous Formulations for Bethe Salpeter Equation
and Radiative Transitions between Two Bound-States}
\author{Chao-Hsi Chang$^{1,2,3}$, Jiao-Kai Chen$^3$ and Guo-Li Wang$^{3,4,5}$}
\address{$^1$ Institute for Advanced Study, Princeton, New Jersey 08540.\\
$^2$CCAST (World Laboratory), P.O.Box 8730, Beijing 100080,
China.\footnote{Not correspondence address.}\\
$^3$Institute of Theoretical Physics, Chinese Academy of Sciences,
P.O. Box 2735, Beijing 100080, China. \\
$^4$ Department of Physics, NanKai University, TianJin 300071, China.\\
$^5$ Department of physics, FuJian Normal University, FuZhou
350007, China.}

\begin{abstract}
The instantaneous formulations for the relativistic Bethe-Salpeter
(BS) and the radiative transitions between the bound-states are
achieved if the BS kernel is instantaneous. It is shown that the
original Salpeter instantaneous equation set up on the BS equation
with an instantaneous kernel should be extended to involve the
`small (negative energy) component' of the BS wave functions. As a
precise example of the extension for the bound states with one
kind of quantum number, the way to reduce the novel extended
instantaneous equation is presented. How to guarantee the gauge
invariance for the radiative transitions which is formulated in
terms of BS wave functions, especially, which is formulated in the
instantaneous formulation, is shown. It is also shown that to
`guarantee' the gauge invariance for the radiative transitions in
instantaneous formulation, the novel instantaneous equation for
the bound states plays a very important role. Prospects on the
applications and consequences of the obtained instantaneous
formulations are outlined.\\



\end{abstract}

\maketitle


\section{Intruction}

It is well-known that Bethe-Salpeter (BS) equation is a
relativistic integration equation in four dimensional Minkowski
space in quantum field theories and it is used to describe bound
states. It was established by analyzing the poles and
corresponding residues of the relevant four-leg Green functions
many years ago\cite{BS}. Thus BS equation has a solid foundation
on quantum field theories. The integration kernel of a BS equation
(the binding interaction for the bound states) corresponding to
the relevant Green functions, in principle, contains an infinite
series of Feynman diagrams, whereas when establishing a precise BS
equation, certain truncation on the infinite series of the Feynman
diagrams must be made by assumption(s) i.e. the kernel just
corresponds one or a few of Feynman diagrams. Besides computing
the spectrum of binding systems, furthermore to formulate the
transitions between the bound states with corresponding BS wave
functions was also achieved\cite{mand,compos}, so that the
applications of BS equation have been extended widely.

Salpeter is the first author, who establishes the precise relation
between four-dimensional BS equations and three-dimensional
Schr\"odinger equations in describing the bound states by means of
the so-called instantaneous approximation\cite{salp} when the
kernel of BS equation is `instantaneous'. In fact, such as
positronium, muonium and atoms etc for instance, are
electromagnetic binding systems, and Coulomb gauge commonly is
known to be far superior for low momentum exchange to describe the
binding of the systems, whereas the leading term in the gauge for
the kernel is exact instantaneous. Thus if one uses BS equation
and the kernel is truncated up to leading term only, then they are
good examples of the kernel being instantaneous exactly. Since
there are more mature methods to solve a Schr\"odinger equation
than that to solve a four dimensional BS equation, so the
so-called instantaneous approximation has been adopted widely in
various applications of BS equations. Hence in literature, when
the kernel indeed is instantaneous as realized by Salpeter, quite
frequently the instantaneous approximation on the BS equation was
made and the obtained equation with certain `corrections' to the
Schrodinger one generally is called as Salpeter equation. Since
the original heavy-quark potential model\cite{Conn} is on
non-relativistic Schr\"odinger equation in three dimension, so
people would like to set the potential model on a more solid
foundation as on a ground of quantum field theory i.e. based on
Salpeter achievement to put the model on a relevant BS equation
with an instantaneous kernel, thus the Salpeter equation has also
received additional attentions.

In fact, heavy-quark potential model itself has made a lot of
phenomenological achievements since it was proposed. Effective
theories, such as heavy quark effective theory (HQET)\cite{hqet},
non-relativistic quantum electromagnetic dynamics
(NRQED)\cite{nrqed} and non-relativistic quantum chromodynamics
(NRQCD)\cite{braat} etc, were established later than potential
model, but they have also achieved wide successes too. Whereas in
the framework of effective theories, when one or more bound states
(mesons) are involved in the considered processes such as decays
and production, they are treated finally to relate to certain
local operator(s) being sandwiched by the bound states with proper
coefficient(s). While the coefficients, being of perturbative
nature, may be calculated by matching them with those of the
underlying quantum field theory, e.g., the full QCD in the cases
for HQET and NRQCD at a suitable high energy scale, whereas the
matrix elements i.e. the operator(s) being sandwiched by the bound
states, being of the non-perturbative nature, can be determined by
fitting experimental data phenomenologically or by lattice QCD
etc, such non-perturbative computations directly. The
non-perterbative computations are difficult and available for
limited cases although they are in progress as the computer power
is increasing. Alternatively, sometimes phenomenological models
may achieve quite accurate results in the estimates of the decays
and production processes. The heavy-quark potential model,
inspired by QCD, is one of them and 'quite powerful'. Thus the
potential model is still used as a 'tool' to calculate decays and
production quite often. With it, one not only calculates spectrum
of the binding system and the `static' matrix elements but also
the matrices for transitions. In this `direction', the potential
model has been applied and tested quite widely in dealing with the
non-perturbative nature of the combining constituents (quarks)
into bound states (mesons) and the transitions. It indeed has
obtained quite a lot of satisfied results. Therefore, to put the
phenomenological potential model on a solid ground of quantum
field theory has been an interesting topic for quite long time. In
fact, in the framework of quantum field theories themselves, the
problems were solved in many years ago that the bound states are
described in terms of BS equation and the transitions between the
bound states are formulated with BS equation solutions (wave
functions)\cite{mand,compos}, so the `shortest' way is e.g. to
embed the potential model for heavy quarks into the BS framework
on quantum field theories QCD.

There was an important progress after Refs.\cite{mand,compos}. It
is about Abel gauge invariance on radiative transition
formulation\cite{lee,changh}. Namely, the so-called irreducible
diagrams appearing in the formulation of the transition matrix
elements\cite{mand,compos} should be determined in certain way by
the BS equation kernel accordingly: the so-called irreducible
diagrams for the transitions, which are truncated from a series of
the relevant Feynman diagrams, should match to the truncation for
the kernel of the BS equation exactly.

For a fundamental process in quantum field theories, the gauge
invariance plays a very important role, and it is guaranteed
perturbatively in Alelian and Non-Abelian cases. In non-Abelian
cases the gauge invariance is much more complicated when bound
states are involved in processes. Here as the first step we
restrict ourselves to focus the Abelian cases in a moment. To
guarantee the gauge invariance, although with bound states being
involved, to work out the precise relation between the irreducible
Feynman diagrams for transitions and those for BS kernel is not
too complicated. In Abelian cases, it is known that for a
fundamental process the problem is so simple that the gauge
invariance is guaranteed, as long as all of the Feynman diagrams
(either tree or loop ones) in a given order are taken into
account. But for a process with one or more bound states being
involved, to guarantee the gauge invariance is not so
straightforward, even one may `trace' them to relevant fundamental
processes accordingly. It is due to the fact, as indicated by BS
equation, that to form a bound state means an infinite series of
certain Feynman diagrams selected by BS kernel are summed, and in
the meantime a lot of diagrams out of the selection of BS kernel
(there are always some in each order) are dropped, therefore,
without careful treatment, the gauge invariance will be lost. The
authors of Ref.\cite{lee} solved the problem for certain cases
first, and the authors of Ref.\cite{changh} generalized it into
general cases. Since Ref.\cite{changh} is in Chinese, so we will
repeat the key points of Ref.\cite{changh} briefly in a suitable
place in this paper.

When considering the decays, such as the decays $B_c \to
J/\psi+\cdots$ etc, and there may be a great (even relativistic)
momentum recoil in the decays due to the mass difference
$m_{B_c}\gg m_{J/\psi}$ (here $J/\psi$ is a bound state of a pair
of charmed quark $c$ and antiquark $\bar{c}$ and $B_c$ is the
ground state of $c$ and $\bar{b}$ guarks), then it is needed to
invent a method to dictate the recoil effects in heavy-quark
potential model properly. To take into account the recoil effects
properly, in Refs.\cite{dec,inst}, the so-called generalized
instantaneous approximation was proposed in terms of BS wave
functions and the relevant formulation for the decays. The
so-called generalized instantaneous approximation was to start
with the formulation in Refs.\cite{mand,compos} for the decays
first, which is relativistic so the recoil effects can be treated
properly no matter how great they are; then it was extended with
further approximations to take a generalized instantaneous
approximation on the formulation for the transition matrix
(amplitude) in whole. Finally a formula for the decays was
obtained, which may dictate the recoil effects well but only
Schr\"dinger wave functions appear\cite{mand,compos}. Whereas
later on when we tried to apply the generalized instantaneous
approximation to an electromagnetic transition with the
achievements, we found the gauge invariance was lost under the
approximation.

In order to solve the fresh problem, i.e., to `recover' the gauge
invariance, we re-examined the generalized instantaneous
approximation and the original Salpeter equation as well, and have
found that some un-necessary approximations were made when
Salpeter derived the equation and when the authors of
Refs.\cite{dec,inst} made the generalized instantaneous
approximation on the transition formulation respectively. We
re-derived the equation without the further approximations, and a
novel instantaneous version of the equation (a group of coupled
equations) to describe the bound states and the formulation for
the transitions between the bound-states are achieved, as long as
the kernel of the BS equation is instantaneous. The instantaneous
formulation for the transitions is related to the novel
instantaneous equation directly. In the paper, after some
outlining the necessary
results\cite{mand,compos,lee,changh,dec,inst} briefly, we present
the re-derivation on the novel instantaneous equations, the `exact
instantaneous formulation' for radiative transitions and specially
how to guarantee the gauge invariance for radiative transitions.
Namely we precisely check the gauge invariance of the novel
instantaneous formulation. In the procedure, one may see that
although  the BS kernel is instantaneous as the start point, the
so-called `small (negative energy) components' of the
instantaneous BS equation (the novel Salpeter equation) are
important to guarantee the gauge invariance, and cannot be dropped
simply.

This paper is organized as follows: in Section II, we take an
example: a fermion and an anti-fermion (a mason is the case),
briefly to review the BS equation, the four dimensional
formulation of a radiative transition\cite{mand,compos} and to
show how to keep gauge invariance for them\cite{lee,changh}. In
Section III, we take the same example re-derive Salpeter equation
from BS equation but with less approximation i.e. to keep all of
the components including the so-called small components (but not
as done by Salpeter to drop the small components). Especially, for
later usages, we also convert the instantaneous formulation in
C.M.S. of the bound state into a `covariant' instantaneous
formulation. Then we present the brief derivation and final
results on instantaneous formulation for the radiative transitions
between bound states from the four dimensional one to effective
three dimensional one. We also check the gauge invariance for the
instantaneous formulation in terms of the four dimensional one. In
Section IV, we discuss the obtained formulations and their
consequences for effective theories etc when the BS equation has
an instantaneous kernel. We make prospects on the applications of
the instantaneous formulations. In Appendices we present some
formulae used in the derivation.

\section{Four Dimensional BS-Formulation for Transitions
between Bound States and Its Gauge Invariance}

In this section, we show the four dimensional formulation for
covariant BS equation and possible radiative transitions between
the bound states in terms of the BS functions and show how to
guarantee the gauge invariance for it etc.

\subsection{The BS Equation}
In general, for a binding system of a fermion and an anti-fermion,
the BS wave functions $\chi_{P\xi}(x_1,x_2)$ and
$\bar{\chi}_{P\xi}(x_1,x_2)$ are defined
\begin{eqnarray}
\label{eq2}
&\chi_{P\xi}(x_1,x_2)=<0|T[\psi(x_1)\bar{\psi}(x_2)]|P\xi>=
e^{iPX}\chi_{P\xi}(x)\,,\nonumber\\
&\chi_{P\xi}(x_1,x_2)=<P\xi|T[\bar{\psi}(x_2)\psi(x_1)]|0>=
e^{-iPX}\bar{\chi}_{P\xi}(x)\,,\\
&X=\alpha_1x_1+\alpha_2x_2\,,\;\;\;\; x=x_1-x_2\,,\;\;\;\;
\alpha_1\equiv \frac{m_2}{m_1+m_2}\,,\;\;\;\;\alpha_2\equiv
\frac{m_1}{m_1+m_2}\,,\nonumber
\end{eqnarray}
where $X, P$ are the C.M. coordinator and the four momentum of the
bound state; $m_1, m_2, x_1, x_2$ are the masses and coordinates
of the $1$-quark ($q_1$) and the $2$-antiquark ($\bar{q_2}$)
respectively. The wave functions in momentum space
\begin{eqnarray}\label{eq22}
&\frac{1}{\sqrt{2\omega}}\chi_{P\xi}(q)=\int d^4x
e^{-iqx}\chi_{P\xi}(x),\;\;\;\frac{1}{\sqrt{2\omega}}
\bar{\chi}_{P\xi}(q)=\int
d^4x e^{iqx}\chi_{P\xi}(x)\,,\nonumber \\
&P=p_1+p_2 \,,\;\;\;\;\; q=\alpha_2p_1-\alpha_1p_2\,,
\end{eqnarray}
where $q, \omega$ are the relative four momentum in the bound
state and the energy of the bound state respectively. In
Eqs.(\ref{eq2},\ref{eq22}) $\xi$ denotes the possible and `inner'
degrees of the freedom (quantum numbers) of the bound state (such
as the mesons $A$ and $B$ etc). Since $\xi$ does not affect our
derivations and theorems in the paper, so we drop it out in all of
the statements later on.

The BS equation generally may be written in momentum space
as\footnote{The derivation of the BS equation can be found in text
book and Ref.\cite{BS}, thus we would not repeat it here.}
\begin{eqnarray}
\label{eqBS} \int{d^4q}[\overrightarrow{O}(P,q',q)-
\overrightarrow{V}(P,q',q)]\chi_{P}(q)=0\,,\nonumber\\
\int{d^4q'}\bar{\chi}_{P}(q')[\overleftarrow{O}(P,q',q)-
\overleftarrow{V}(P,q',q)]=0\,.
\end{eqnarray}
The `differential' operators $\overrightarrow{O}$ and
$\overleftarrow{O}$ here acting on the BS wave function under
integration accordingly are defined as the follows:
\begin{eqnarray}
\label{eqBS11}
\int d^4 q
\overrightarrow{O}(P,p,q)\chi_{P}(q)=\int d^4 q
\delta^4(p-q)[S^{(1)}_f(\alpha_1P+p)]^{-1}
\chi_{P}(q)[S^{(2)}_f(-\alpha_2P+q)]^{-1}\,,\nonumber \\
\int d^4 p \bar{\chi}_{P}(p)\overleftarrow{O}(P,p,q)= \int d^4 p
\delta^4(p-q)[S^{(2)}_f(-\alpha_2P+q)]^{-1}
\bar{\chi}_{P}(p)[S^{(1)}_f(\alpha_1P+p)]^{-1}\,.
\end{eqnarray}
To be explicit, here we use $S_f$ with the superscripts ``${(1)}$"
and ``${(2)}$" to denote the propagators of the quark $q_1$ and
the anti-quark $\bar{q}_2$ respectively. Throughout the paper even
without explicitly explaining, we will use the superscripts to
denote the relevant quantities or operators, i.e., to denote those
for the quark $q_1$ and the anti-quark $\bar{q}_2$ respectively as
here. While $\overrightarrow{V}(P,q',q)$ and
$\overleftarrow{V}(P,q'q)$ are binding interaction i.e. the BS
kernel. In addition, the BS wave function $\chi_{P}(q)$ satisfies
the normalization condition:
\begin{equation}
\label{eqnorm} \int\int\frac{d^{4}q d^{4}q'}
{(2\pi)^{4}}Tr\left\{\overline\chi_{P}(q)
\frac{\partial}{\partial{P_{0}}}\left([S_f^{-1}(p_{1})]^{(1)}
[S_f^{(2)}(-p_{2})]^{-1}\delta^{4}(q-q')+
V(p,q,q')\right)\chi_{p}(q')\right\}=2iP_{0}\,,
\end{equation}
here $P_0=\omega$ is the zero component of $P$.

If the $\delta$ function in Eq.(\ref{eqBS}), contained in operator
$\overrightarrow{O}(P,p,q)$ as indicated in Eq,(\ref{eqBS11}), is
just integrated out, the Bethe-Salpeter (BS) equation
corresponding a quark-antiquark bound state can be written as:

\begin{equation}\label{bs00}
(\not\!{p_{1}}-m_{1})\chi_{P}(q)(\not\!{p_{2}}+m_{2})=
\int\frac{d^{4}k}{(2\pi)^{4}}V(P,q,k)\chi_{P}(k)\,,
\end{equation}
or
\begin{equation}\label{bs01}
\chi_{P}(q)=\frac{1}{\not\!{p_{1}}-m_{1}}
\int\frac{d^{4}k}{(2\pi)^{4}} V(P,q,k)
\chi_{P}(k)\frac{1}{\not\!{p_{2}}+m_{2}}\,,
\end{equation}
which is independent on the specific reference frame because it is
Lorentz invariant.

\subsection{The Formulation for Radiative Transitions}

\begin{figure}
\centering
\hfill\includegraphics[width=0.42\textwidth]{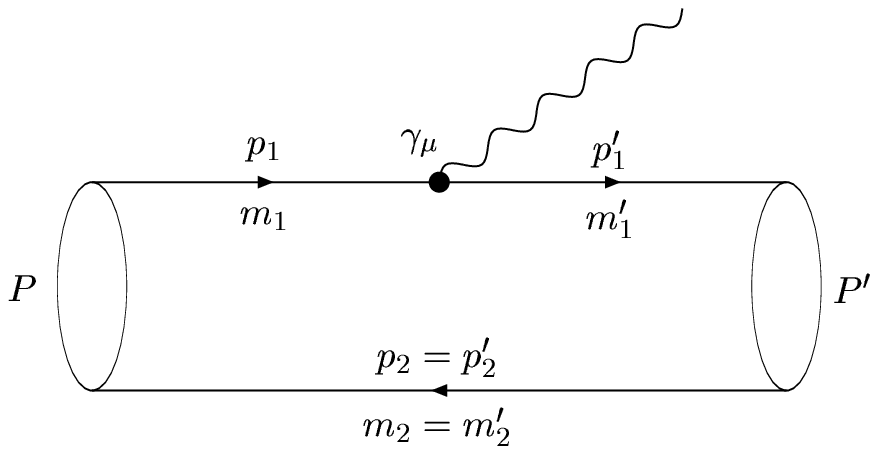}%
\hspace*{\fill} \caption{Feynman diagram corresponding to the
quark in the `meson' $A (B)$ emitting the photon for the
transition $A(P)\to B(P')+\gamma(Q)$.} \label{fig1}
\end{figure}
\begin{figure}
\centering
\hfill\includegraphics[width=0.42\textwidth]{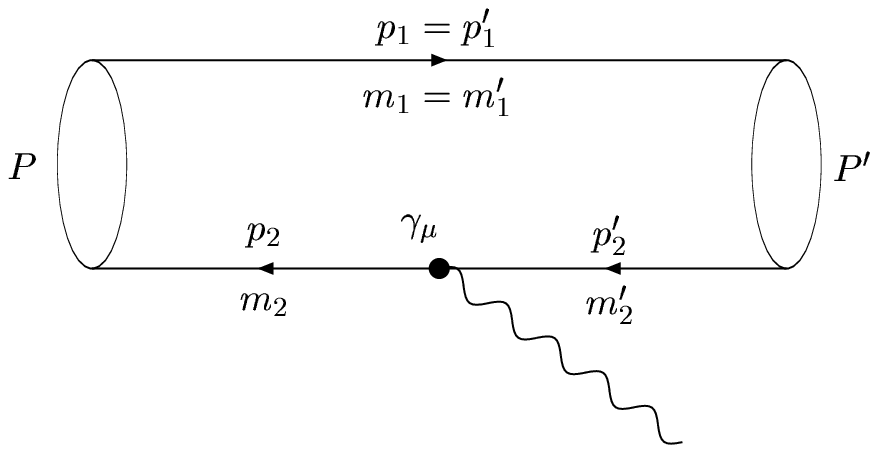}%
\hspace*{\fill} \caption{Feynman diagram corresponding to the
anti-quark in the `meson' $A (B)$ emitting the photon for the
transition $A(P)\to B(P')+\gamma(Q)$.} \label{fig2}
\end{figure}

As an example of the radiative transitions, let us consider a
decay $A(P)\to B(P')+\gamma(Q)$, and we will take this example to
show the formulation for the decays in terms of the BS wave
functions and how the electromagnetic (an Abelien) gauge
invariance is guaranteed throughout the paper. Here the `meson'
$A$ (a bound state) consists of a quark $q_1: p_1, m_1$ and an
anti-quark $\bar{q}_2: p_2, m_2$, and the `meson' $B$ (another
bound state), similar to $A$, consists of a quark $q'_1: p'_1,
m'_1$ and an anti-quark $\bar{q}'_2: p'_2, m'_2$. To be
responsible for the decay, the Feynman diagrams for the lowest
order are described as Fig.\ref{fig1} and Fig.\ref{fig2}.

Since the concerned process is an electromagnetic transition, and
if it is `automatic' decay, the meson $A$ must be an excitation
state of the meson $B$, i.e., we must have $m_1=m'_1, m_2=m'_2$.
For more specific, we further assume that in the mesons, the
quarks $q_1, q'_1: (p_1, m_1), (p'_1, m'_1)$, carry charge $Q_1$
in the unit $e$ and the anti-quarks $\bar{q}_2, \bar{q}'_2: (p_2,
m_2), (p'_2, m'_2)$, carry charge $Q_2$ in the unit $e$. In fact,
in many discussions, proving the gauge invariance and the other
derivations, the contributions from quark $q_1$ Fig.\ref{fig1} and
those from anti-quark $\bar{q_2}$ Fig.\ref{fig2} are independent,
thus to investigate the decay amplitude, either the quark emission
of a photon Fig.\ref{fig1} or the anti-quark emission
Fig.\ref{fig2} may be taken as `a representation' of the
transition (to consider one of them is enough, the other is almost
the same just to repeat once more if one consider both of them for
completeness, so here we say `a representation'). In
Refs.\cite{mand,compos}, how to write down the S-matrix element
for this kind of decays in terms of BS wave function is given:
$$<P',Q\,\epsilon|S|P>=\frac{(2\pi)^4e}{\sqrt{2^3\omega\omega_A\omega_B}}
\delta^4(P'+Q-P)\epsilon^\mu \cdot <P'|J_\mu|P>,$$ here
$\epsilon^\mu$ is the polarization vector of the photon and the
matrix element for the current (Figs.1,2):
\begin{eqnarray}
\label{eq1} &&<P'|J_\mu|P>=\frac{1}{(2\pi)^{4}}\int d^4q' \int
d^4q Tr\left\{\bar{\chi}_{P'}(q') \left[
G^{(1)}_\mu(P'q';Q;P,q)+G^{(2)}_\mu(P'q';Q;P,q)\right]{\chi}_{P}(q)
\right\}\nonumber\\
&&=\frac{1}{(2\pi)^{4}}\int d^4q' \int d^4q
Tr\left\{\bar{\chi}_{P'}(q') Q_1\gamma^{(1)}_{\mu}
{\chi}_{P}(q)[S^{(2)}_{f}(q'-\alpha'_2P')]^{-1} \cdot
\delta^{4}(q-q'+\alpha'_2P'-\alpha_2P)
\right.\nonumber\\
&&\left. + \bar{\chi}_{P'}(q') [S^{(1)}_{f}(\alpha_1P+q)]^{-1}
{\chi}_{P}(q) Q_2\gamma^{(2)}_{\mu} \cdot
\delta^{4}(q-q'+\alpha_1P-\alpha'_1P')\right\}\,,
\end{eqnarray}
namely corresponding to the irreducible Feynman diagrams
Figs.(\ref{fig1},\ref{fig2}), the operators (factors)
$G^{(i)}_\mu(P'q';Q;P,q), (i=1,2)$ are defined by\footnote{Since
the operator (factor) makes sense only when it `acts' on BS wave
functions so here its definition is given always with the wave
functions properly. We note that throughout the paper there are
several operators which are defined in the same manner as here.}
$$\bar{\chi}_{P'}(q')G^{(1)}_\mu(P'q';Q;P,q){\chi}_{P}(q)\equiv
\bar{\chi}_{P'}(q') Q_1\gamma^{(1)}_\mu{\chi}_{P}(q)
[S^{(2)}_f(q'-\alpha'_2P')]^{-1}\delta(q-q'+\alpha'_2P'-\alpha_2P)\,;$$
$$\bar{\chi}_{P'}(q')G^{(2)}_\mu(P'q';Q;P,q){\chi}_{P}(q)\equiv
\bar{\chi}_{P'}(q') [S^{(1)}_{f}(\alpha_1P+q)]^{-1} {\chi}_{P}(q)
Q_2\gamma^{(2)}_{\mu} \cdot
\delta^{4}(q-q'+\alpha_1P-\alpha'_1P')\,.$$ Here the quarks'
momenta, $p_1, p'_1$, and anti-quarks' momenta, $p_2, p'_2$, are
related to the total momenta $P, P'$ and the relative ones $q, q'$
of the initial meson $A$ and the finial meson $B$ respectively:
\begin{eqnarray}
\label{eqqp} p_{1}={\alpha}_{1}P+q\,, \;\; {\alpha}_{1}\equiv
\frac{m_{1}}{m_{1}+m_{2}}\,,\;\;\; p_{2}={\alpha}_{2}P-q, \;\;
{\alpha}_{2}\equiv \frac{m_{2}}{m_{1}+m_{2}}\,;\nonumber \\
p'_{1}={\alpha}'_{1}P'+q'\,, \;\; {\alpha}'_{1}\equiv
\frac{m'_{1}}{m'_{1}+m'_{2}}\,,\;\;\; p'_{2}={\alpha}'_{2}P'-q'\,,
\;\; {\alpha}'_{2}\equiv \frac{m'_{2}}{m'_{1}+m'_{2}}\,.
\end{eqnarray}
In Eq.(\ref{eq1}) `$Tr$' denotes the trace for the
$\gamma$-matrices is taken. ${\chi}_{P}(q)$ and
$\bar{\chi}_{P'}(q')$ are the relevant BS wave functions of the
mesons $A$ and $B$; $S^{(i)}_f (i=1,2)$ are the propagator of the
quark and anti-quark. $Q=P-P'$ is the photon's four momentum.
Since the decay is electromagnetic, i.e., the flavor(s) is(are)
conserved, so we have $m_1=m'_1$ and $m_2=m'_2$, that
$\alpha_1=\alpha'_1, \alpha_2=\alpha'_2$ and
$\alpha'_iP'-\alpha_iP=-\alpha_iQ, (i=1,2)$. Thus later on we
simplify Eq.(\ref{eq1}) by taking $m_1=m'_1$ and $m_2=m'_2$
always.

In Refs.\cite{mand,compos}, the operators
$G^{(1)}_\mu(P'q';Q;P,q)$ and $G^{(2)}_\mu(P'q';Q;P,q)$ are
obtained by analyzing the Feynman diagrams and to be sure only the
`irreducible' diagrams are involved. In fact, the amplitude being
gauge invariant also means the operators have very definite
relation to the BS equation (the `differential operator
$\overrightarrow{O}(P,q,q')$ or $\overleftarrow{O}(P,q,q')$' and
the kernel $V(P,q,q')$). Let us show the relation of the operators
to the BS equation in the following subsection.

\subsection{The Electromagnetic Gauge Invariance of the
Radiative Transitions}

According to Refs.\cite{lee,changh}, when a transition current
matrix element, in which there is(are) bound state(s) being
involved, is formulated in terms of BS equation, such as
Eq.(\ref{eq1}), and if one requests it is gauge invariant (the
current is conserved), then the transition operators
$G^{(i)}_\mu(P',q';Q;P,q) (i=1,2)$, which appear in Eq.(\ref{eq1})
and correspond to the irreducible Feynman diagrams\footnote{In
fact, the request only, that `the operators
$G^{(i)}_\mu(P',q';Q;P,q)\; (i=1,2)$ correspond to irreducible
Feynman diagrams', cannot determine the operators completely.},
must `match' with the BS equation accordingly in an exact
way\footnote{Since the operators in BS equation also correspond to
irreducible Feynman diagrams, so there is no conflict with the
irreducible request on the transition operators
$G^{(i)}_\mu(P',q';Q;P,q)\; (i=1,2)$. Whereas with the match
condition the transition operators will be well determined.}.
Therefore, to determine a specific $G^{(i)}_\mu(P',q';Q;P,q)$ and
to guarantee the matrix element being gauge invariant, we need to
explore the match rule precisely.

Now let us `explore' the match rule and show the specific
$G^{(i)}_\mu(P',q';Q;P,q)$ happens to be satisfied with the match
rule, when the BS wave functions adopted in Eq.(\ref{eq1})
correspond to the solutions of the equation with the `lowest'
order kernel.

To explore the `match rule', first of all, let us introduce an
external classical field $A^{ex}_\mu$ and in the underlying theory
add a coupling term:
\begin{equation}
\label{eq0} J^{\mu}_{em}(x)A^{ex}_{\mu}(x),
\end{equation}
to the Lagrangian, here $J^\mu_{em}$ is the electromagnetic
current of the field theory. Then being parallel the case without
the $A^{ex}_\mu$, one now may establish a BS equation with the
external field $A^{ex}_\mu$ through the coupling term
Eq.(\ref{eq0}), and the BS equation operators
$\overrightarrow{O^{A^{ex}_\mu}}$ and
$\overrightarrow{V^{A^{ex}_\mu}}$ may be obtained accordingly.
With the preparations, the `match rule' can be stated as that
$G_\mu$, the irreducible Feynman diagrams, should be exactly
obtained as the the following way:
\begin{equation}
\label{eqA}
\{G_\mu\}\equiv\sum_{i=1,2}G^{(i)}_\mu(P',q';Q;P,q)=
\frac{\delta(\overrightarrow{O^{A^{ex}_\mu}}-
\overrightarrow{V^{A^{ex}_\mu}})}{\delta A^{ex}_\mu}\;
|_{A^{ex}_\mu\rightarrow 0}\,.
\end{equation}
The Eq.(\ref{eqA}), in fact, can be explained as that the
irreducible Feynman diagrams, $G_\mu$, should be obtained by the
way that the photon line attaches onto each charged line in the BS
equation operators $O(P,q',q)-V(P,q',q)$ once in turn. To be
precise, let us take the example to explain Eq.(\ref{eqA}) and its
meaning.

With the match rule between $G_\mu(P',q';Q;P,q)$ and BS equation,
let us turn back to examine the specific
$G^{(i)}_\mu(P',q';Q;P,q)$ in Eq.(\ref{eq1}). If the BS kernel
$V(P,q',q)$ for the bound states $P,P'$ of quark and anti-quark is
just one gluon exchange
$$\int d^4q' \bar{\chi}_P(q')V(P,q',q)=\int d^4q' \gamma_\mu^{(2)}
\bar{\chi}_P(q')\gamma_\mu^{(1)} D_f(q'-q)$$ or
\begin{eqnarray}
\label{eqker} \int d^4q V(P,q',q)\chi_P(q)=\int d^4 q
\gamma_\mu^{(1)} \chi_P(q)\gamma_\mu^{(2)} D_f(q'-q)\,,
\end{eqnarray}
i.e. $D_f$ is the gluon propagator, and in the underlying theory
the quark carries charge $Q_1$ in unit of $e$ and the anti-quark
carries $Q_2$ that means in the electromagnetic current
$J^\mu_{em}$ Eq.(\ref{eq0}) there are quark current
$Q_1J^{(1)}_\mu$ and anti-quark $Q_2J^{(2)}_\mu$, hence by
Eq.(\ref{eqA}) for the transition the irreducible Feynman diagrams
$G^{(1)}_\mu(P',q';Q;P,q)$ is generated by attaching the photon
line onto the quark line in $O(P,q',q)-V(P,q',q)$ (Fig.1a) and
$G^{(2)}_\mu(P',q';Q;P,q)$ is generated by attaching the photon
line onto the anti-quark line in $O(P,q',q)-V(P,q',q)$ (Fig.1b) as
indicated in Eq.(\ref{eq1}). Note here that since $V(P,q',q)$ is
neutral so according to the match rule there is no contribution
from $V(P,q',q)$ at all, thus the photon line can attach onto
those lines in $O(P,q',q)$ of the BS equation only.

Now following Refs.\cite{lee,changh}, let us show the gauge
invariance, i.e. the current matrix element is conserved.

To apply the equations $Q=P-P'$ and the general identity
\begin{eqnarray}
-iQ^{\mu}\gamma^{(1)}_{\mu}=
[S^{(1)}_{f}(\alpha_1P+q)]^{-1}
-[S^{(1)}_{f}(\alpha'_1P'+q')]^{-1}\,,\nonumber\\
iQ^{\mu}\gamma^{(2)}_{\mu}= [S^{(2)}_{f}(-\alpha_2
P+q)]^{-1}-[S_{f}^{(2)}(-\alpha'_2P'+q')]^{-1}\,,
\end{eqnarray}
we have
\begin{eqnarray}
\label{eqcurr0} &-iQ^{\mu}<P'|J_\mu|P>= \int d^4q'\int
d^4qTr\left\{\bar{\chi}_{P'}(q')
Q_1\left([S_{f}^{(1)}(\alpha_1P+q)]^{-1}
\right.\right. \nonumber \\
&\left.\left.-[S_{f}^{(1)}(\alpha'_1P'+q')]^{-1}\right)
{\chi}_{P}(q)[S^{(2)}_{f}(-\alpha'_2P'+q')]^{-1}
\delta^{4}(q-q'-\alpha_2Q)-\right. \nonumber \\
&\left. \bar{\chi}_{P'}(q')[S_{f}^{(1)}(\alpha_1P+q)]^{-1}
{\chi}_{P}(q)Q_2
\left([S^{(2)}_{f}(-\alpha_2P+q)]^{-1}
-[S_{f}^{(2)}(-\alpha'_2P'+q')]^{-1}
\right)
\delta^{4}(q-q'+\alpha_1Q)\right\}\nonumber \\
&=\int d^4q'\int
d^4qTr\bar{\chi}_{P'}(q')\left\{Q_1\left[\overrightarrow{O}(P,q'+\alpha_2Q,q)
-\overleftarrow{O}(P',q',q-\alpha_2Q)\right]+
Q_2\left[\cdots\right]\right\}{\chi}_{P}(q)\,.
\end{eqnarray}
Considering the property of the BS kernel $V$ indicated by Eq.(\ref{eqker})
$$\bar{\chi}_{P'}(q')\overrightarrow{V}(P,q'+\alpha_2Q,q)\chi_{P}(q)
=\bar{\chi}_{P'}(q')\overleftarrow{V}(P',q',q-\alpha_2Q)\chi_{P}(q)\,,$$
and
\begin{eqnarray}
\bar{\chi}_{P'}(q')\overrightarrow{V}(P,q'-\alpha_1Q,q)\chi_{P}(q)
=\bar{\chi}_{P'}(q')\overleftarrow{V}(P',q',q+\alpha_1Q)\chi_{P}(q)\,,
\end{eqnarray}
then Eq.(\ref{eqcurr0}) can be rewritten
\begin{eqnarray}
\label{eqcurr1} &
-iQ^{\mu}<P'|J_\mu|P>= \nonumber\\
&\int d^4q'\int d^4qTr\left\{
Q_1\bar{\chi}_{P'}(q')\left[\overrightarrow{O}(P,q'+\alpha_2Q,q)
-\overrightarrow{V}(P,q'+\alpha_2Q,q)\right]{\chi}_{P}(q) \right. \nonumber \\
&-\left.Q_1\bar{\chi}_{P'}(q')\left[\overleftarrow{O}(P',q',q-\alpha_2Q)
-\overleftarrow{V}(P',q',q-\alpha_2Q)\right]{\chi}_{P}(q)\right.
\nonumber\\
&\left. +
Q_2\bar{\chi}_{P'}(q')\left[\overrightarrow{O}(P,q'-\alpha_1Q,q)
-\overrightarrow{V}(P,q'-\alpha_1Q,q)\right]{\chi}_{P}(q) \right. \nonumber \\
&-\left.Q_2\bar{\chi}_{P'}(q')\left[\overleftarrow{O}(P',q',q+\alpha_1Q)
-\overleftarrow{V}(P',q',q+\alpha_1Q)\right]{\chi}_{P}(q)\right\}
\,.
\end{eqnarray}
It is easy to see when the BS equations Eq.(\ref{eqBS}) are
applied to Eq.(\ref{eqcurr1}) for $\bar{\chi}_{P'}(p)$ and
$\chi_P(q)$ respectively, the matrix element of the current is
conserved
\begin{eqnarray}
\label{eqcurr2} &-iQ^{\mu}<P'|J_\mu|P>=0\,,
\end{eqnarray}
i.e. the Abel gauge invariance is proved.

Since the charge values $Q_1$ and $Q_2$ of the quark and
anti-quark are independent, and one may also see from the above
that the terms proportional to $Q_1$ and $Q_2$ are independent
too, thus to prove the gauge invariance, it is enough to show one
case, for instance, for the terms which are proportional to $Q_1$
only. Namely it is enough to show the terms proportional to $Q_1$
themselves are gauge invariance (for the terms which are
proportional to $Q_2$ to show their gauge invariance is similar).
Later on to shorten the presentation of the derivation, we only
present those terms either proportional to $Q_1$ or to $Q_2$
precisely, and just add the statement as `the other terms are
similar' for instance.

\section{The Instantaneous Formulation for Radiative Transitions
between Bound States and Its Gauge Invariance}

Before deriving the instantaneous formulation for the transitions
between bound states, first of all, we need to do some
preparations, namely we need to re-write the `instantaneous' BS
equation in covariant formulation in a reference frame, in which
the bound state may be moving.

\subsection{The BS Equation with an Instantaneous Kernel and Its
reduced Equation}

The so-called `instantaneous approximation' was realized by
Salpeter first, and it is not Lorentz covariant\cite{salp}. In his
original paper and in most literature, indeed an approximation was
made in the rest frame of the concerned bound state. It says when
the BS has an instantaneous kernel i.e. the kernel of BS does not
depend on the time component of the relative momentum $k_0$, such
as to take a simple form
\begin{eqnarray} \label{eqinst}
V(P^{\mu},k^{\mu},q^{\mu})|_{\overrightarrow{P}=0} \simeq
V(|\stackrel{\rightarrow}{k}- \stackrel{\rightarrow}{q}|)\,,
\end{eqnarray}
then as a main result of the `instantaneous approximation', the
four dimension BS equation can be reduced into a three-dimension
Schr\"odinger equation\cite{salp}.

Whereas here when BS equation really has an instantaneous kernel,
we reduce the BS equation without approximation in certain sense.
Especially, we are considering the radiative transitions between
the bound states, and we find that with an approximation as done
by Salpeter the electromagnetic gauge invariance for the
transitions cannot be guaranteed, thus we need to start the
reducing of the BS equation with an instantaneous kernel from
beginning, but to reduce the `instantaneous equation and the wave
function' in such a way that is the so-called instantaneous but
Lorentz covariant.

To do so, first of all we divide the relative momentum $q$ into
two components, $q_{_\parallel}$ and $q_{_\perp}$, a parallel
component and an orthogonal one to the bound state momentum $P\;\;
(P^2=M^2)$ respectively:
$$q^{\mu}=q^{\mu}_{P_\parallel}+q^{\mu}_{P_\perp},$$
where $q^{\mu}_{P_\parallel}\equiv \frac{(P\cdot q)}{M^{2}}
P^{\mu}$ and $q^{\mu}_{P_\perp}\equiv
q^{\mu}-q^{\mu}_{P_\parallel}$. Correspondingly, we have two
Lorentz invariant variables:
$$q_{P}=\frac{(P\cdot q)}{M}\,,\;\;\;\;
q_{P_T}=\sqrt{q^{2}_{P}-q^{2}}=\sqrt{-q^{2}_{P_\perp}}\,.$$ In the
rest frame of the concerned bound state, i.e.,
$\stackrel{\rightarrow}{P}=0$, they turn to the usual components
$q_{0}$ and $|\overrightarrow{q}|$, respectively. Now the
integration element of the relative momentum $k$ can be written in
an invariant form:
\begin{equation}
d^{4}k=dk_{P}k^{2}_{P_T}dk_{P_T}ds d\phi,
\end{equation}
where $\phi$ is the azimuthal angle, $s=\frac{k_{P}q_{P}-(k\cdot
q)}{k_{P_T}q_{P_T}}$. The interaction kernel Eq.(\ref{eqinst}) can
be re-written as:
\begin{equation}
\label{eqinst1} V(P^{\mu},k^\mu,q^\mu)=
V(|k^{\mu}_{P_\perp}-q^{\mu}_{P_\perp}|).
\end{equation}

Similar to Ref.\cite{salp}, we introduce the `instantaneous' BS
wave function:
\begin{equation}
\label{eqinstw} \varphi_{P}(q^{\mu}_{P_\perp})\equiv i\int
\frac{dq_{P}}{2\pi}\chi_{P}(q^{\mu}_{P_\parallel},q^{\mu}_{P_\perp})\,,
\end{equation}
and integration
\begin{equation}
\eta(q^{\mu}_{P_\perp})\equiv\int\frac{k^{2}_{P_T}dk_{P_T}ds}{(2\pi)^{2}}
V(|k^\mu_{P_\perp}-q^\mu_{P_\perp}|)\varphi_{P}(k^{\mu}_{P_\perp})\,,
\end{equation}
then the BS equation Eq.\ref{bs01} becomes
\begin{equation}
\label{eqBSin}
\chi_{P}(q^\mu_{P_\parallel},q^\mu_{P_\perp})=S_f^{(1)}(p^\mu_{1})
\eta(q^\mu_{P_\perp})S_f^{(2)}(-p^\mu_{2}),
\end{equation}
where $S_f^{(1)}(p_{1})$ and $S_f^{(2)}(-p_{2})$ are the
propagators of the quark and anti-quark respectively. They can be
decomposed as:
\begin{equation}
\label{eqpropo}
-iJS_f^{(i)}(Jp^\mu_{i})=\frac{\Lambda^{+}_{ip}(q^\mu_{P_\perp})}{Jq_{P}
+\alpha_{i}M-\omega_{iP}+i\epsilon}+
\frac{\Lambda^{-}_{ip}(q^\mu_{P_\perp})}{Jq_{P}+\alpha_{i}M+
\omega_{iP}-i\epsilon}\,,
\end{equation}
with
\begin{equation}
\label{eqpropo1}
\omega_{iP}=\sqrt{m_{i}^{2}+q^{2}_{p_T}}\,,\;\;\;\;
\Lambda^{\pm}_{iP}(q_{P_\perp})= \frac{1}{2\omega_{iP}}\left[
\frac{\not\!\!{P}}{M}\omega_{iP}\pm
J(m_{i}+{\not\!q}_{P_\perp})\right],
\end{equation}
where $J=1$ for the quark($i=1$) and $J=-1$ for the
anti-quark($i=2$). Here $\Lambda^{\pm}_{iP}(q_{P_\perp})$
satisfies the relations:
\begin{equation}
\label{eq24}
\Lambda^{+}_{iP}(q^\mu_{P_\perp})+
\Lambda^{-}_{iP}(q^\mu_{P_\perp})=
\frac{\not\!\!{P}}{M}\,,\;\;
\Lambda^{\pm}_{iP}(q^\mu_{P_\perp})\frac{\not\!\!{P}}{M}
\Lambda^{\pm}_{iP}(q^\mu_{P_\perp})
=\Lambda^{\pm}_{iP}(q^\mu_{P_\perp})\,,\;\;
\Lambda^{\pm}_{iP}(q^\mu_{P_\perp})\frac{\not\!\!{P}}{M}
\Lambda^{\mp}_{iP}(q^\mu_{P_\perp})=0\,.
\end{equation}
Namely $\Lambda^{\pm}$ may be considered as projection operators
(the energy ($P$)-projection operators), while in the rest frame
of the bound state, they turn to the energy projection operator.

For later discussions let us define
$\varphi^{\pm\pm}_{P}(q^\mu_{P_\perp})$ as:
\begin{equation} \label{defini}
\varphi^{\pm\pm}_{P}(q^\mu_{P_\perp})\equiv
\Lambda^{\pm}_{1P}(q^\mu_{P_\perp})
\frac{\not\!\!{P}}{M}\varphi_{P}(q^\mu_{P_\perp})
\frac{\not\!\!{P}}{M} \Lambda^{\pm}_{2P}(q^\mu_{P_\perp})\,,
\end{equation}
then for the BS wave functions $\varphi_{P}(q^\mu_{P_\perp})$ we
have:
$$\varphi_{P}(q^\mu_{P_\perp})=\varphi^{++}_{P}(q^\mu_{P_\perp})+
\varphi^{+-}_{P}(q^\mu_{P_\perp})+\varphi^{-+}_{P}(q^\mu_{P_\perp})
+\varphi^{--}_{P}(q^\mu_{P_\perp})\,.$$

Since the BS equation kernel is instantaneous, we may integrate
out $q_{P}$ (contour integration) on both sides of
Eq.(\ref{eqBSin}), and obtain
\begin{eqnarray}
\label{eqcont}
\varphi_{P}(q^\mu_{P_\perp})=\frac{\Lambda^{+}_{1P}(q^\mu_{P_\perp})
\eta_{P}(q^\mu_{P_\perp})\Lambda^{+}_{2P}(q^\mu_{P_\perp})}
{(M-\omega_{1P}-\omega_{2P})}-\frac{\Lambda^{-}_{1P}(q^\mu_{P_\perp})
\eta_{P}(q^\mu_{P_\perp})\Lambda^{-}_{2P}(q^\mu_{P_\perp})}
{(M+\omega_{1P}+\omega_{2P})}\,,
\end{eqnarray}
and applying the energy ($P$)-projection operators
$\Lambda^\pm_{iP}(q^\mu_{P_\perp})$ to Eq.(\ref{eqcont}) further,
we obtain the coupled equations:
\begin{eqnarray}
\label{eqpp}
(M-\omega_{1P}-\omega_{2P})\varphi^{++}_{P}(q^\mu_{P_\perp})=
\Lambda^{+}_{1P}(q^\mu_{P_\perp})
\eta_{P}(q^\mu_{P_\perp})\Lambda^{+}_{2P}(q^\mu_{P_\perp})\,,
\end{eqnarray}
\begin{eqnarray}
\label{eqmm}
(M+\omega_{1P}+\omega_{2P})\varphi^{--}_{P}(q^\mu_{P_\perp})=
-\Lambda^{-}_{1P}(q^\mu_{P_\perp})
\eta_{P}(q^\mu_{P_\perp})\Lambda^{-}_{2P}(q^\mu_{P_\perp})\,,
\end{eqnarray}
\begin{eqnarray}
\label{eqpm}
\varphi^{+-}_{P}(q^\mu_{P_\perp})=\varphi^{-+}_{P}(q^\mu_{P_\perp})=0\,.
\end{eqnarray}

The normalization condition Eq.(\ref{eqnorm}) now terns to the
follows accordingly by the positive and negative functions as:
\begin{equation}
\int\frac{q_T^2dq_T}{(2\pi)^2}tr\left[\overline\varphi^{++}
\frac{{/}\!\!\!
{p}}{M}\varphi^{++}\frac{{/}\!\!\!{p}}{M}-\overline\varphi^{--}
\frac{{/}\!\!\! {p}}{M}\varphi^{--}\frac{{/}\!\!\!
{p}}{M}\right]=2P_0.
\end{equation}

Note that Salpeter and many authors in literature considered that,
of the above coupled equations
Eqs.(\ref{eqpp},\ref{eqmm},\ref{eqpm}), the equation
Eq.(\ref{eqpp}) is `dominant', owing to having a very small
coefficient $(M-\omega_{1P}-\omega_{2P})$ (weak binding) on the
left hand side, and it is `projected' by positive-positive energy
project operators ($\Lambda^{+}_{1P}(q^\mu_{P_\perp})$ and
$\Lambda^{+}_{2P}(q^\mu_{P_\perp})$) (it is why people call it as
positive energy equation). Thus the authors highlighted on it only
and dropped the rest equations out at all. With the approximation,
they showed that if expending the equation Eq.(\ref{eqpp}) to the
order of $\frac{q^2}{M^2}$ further, the equation may be turned
into a Schr\"odinger one accordingly\footnote{In fact, exactly to
say it is misleading. It is because that, only when they made a
further assumption on the spin structure of the BS wave function
properly, the equation Eq.(\ref{eqpp}) may be turned into a
Schr\"odinger equation\cite{salp}.}. Whereas, as emphasized in
Introduction, we realized that if only keeping Eq.(\ref{eqpp}) and
its solutions as wave functions to be applied to the radiative
transitions, the gauge invariance will be violated. Thus here in
the next two subsections we will show the formulation of the
transitions between bound states in four dimension how to turns
into an instantaneous one and accordingly how to keep the gauge
invariance when the formulation in four dimension is gauge
invariant precisely. It is just by considering all the coupled
equations Eqs.(\ref{eqpp},\ref{eqmm},\ref{eqpm}) `equally', and
one will see that all of the equations play important roles in
guaranteeing the gauge invariance for the radiative transitions
between bound states.

To explore the contents of the
Eqs.(\ref{eqpp},\ref{eqmm},\ref{eqpm}), here let us take a heavy
quarkonium (or positronium) as samples to establish the coupled
equations for the components in the instantaneous BS wave
functions $\varphi_{P}(q^\mu_{P_\perp})$ and to show the equations
are consistent.

Since a heavy quarkonium $(Q\bar{Q})$ is of equal masses
($m_1=m_2$, i.e., $\alpha_1=\alpha_2=\frac{1}{2}$), so its
physical states have definite quantum number of $J^{PC}$, thus its
equations should be given according to the quantum number
precisely. Here we take the state with $J^{PC}=0^{-+}$ (total spin
$S=0$, and total orbital angle moment $L=0$) as an example to show
the main feature of the equations
Eqs.(\ref{eqpp},\ref{eqmm},\ref{eqpm}). Of the components, the
general relativistic wave function for the bound state with the
quantum number $J^{PC}=0^{-+}$ (in the center of mass system) can
be written as:
\begin{equation}
\varphi_{^{1}S_0}(\stackrel{\rightarrow}{q})=
\left[{\gamma_0}\varphi
_1(\stackrel{\rightarrow}{q})+\varphi_2(\stackrel{\rightarrow}{q})+
{\not\!q}_{\perp} \varphi_3(\stackrel{\rightarrow}{q})+{\gamma_0}
{\not\!q}_{\perp}
\varphi_4(\stackrel{\rightarrow}{q})\right]\gamma_5\,,
\end{equation}
where ${q}_{\perp}=(0,\stackrel{\rightarrow}{q})$. The equation
Eq.(\ref{eqpm}), in fact it acts as constraints, give the strong
conditions precisely:
\begin{equation}\label{constr0}
\varphi_1(\stackrel{\rightarrow}{q})
=-m_1\varphi_4(\stackrel{\rightarrow}{q}), \;\;\;\;
\varphi_3(\stackrel{\rightarrow}{q})=0\,.
\end{equation}
Then we can rewrite the relativistic wave function of state
$0^{-+}$ as:
\begin{equation}
\varphi_{^{1}S_0}(\stackrel{\rightarrow}{q})=\left[
{\gamma_0}\varphi_1(\stackrel{\rightarrow}{q})
+\varphi_2(\stackrel{\rightarrow}{q})-
{\gamma_0}{\not\!q}_{\perp}\frac{1}
{m_1}\varphi_1(\stackrel{\rightarrow}{q})\right]\gamma_5\,.\label{0-+}
\end{equation}
To apply the definition Eq.(\ref{defini}) to Eq.(\ref{0-+}), the
corresponding positive and negative projected wave functions may
be written as:
\begin{equation}
\varphi^{++}_{^{1}S_0}(\stackrel{\rightarrow}{q})=
\frac{1}{2\omega_1m_1}\left(\omega_1\varphi_1(\stackrel{\rightarrow}{q})
+m_1\varphi_2(\stackrel{\rightarrow}{q})\right)\left({\gamma_0}m_1+
\omega_1-{\gamma_0}{\not\!q}_{\perp}\right)\gamma_5\,,
\end{equation}
and
\begin{equation}
\varphi^{--}_{^{1}S_0}(\stackrel{\rightarrow}{q})=
\frac{1}{2\omega_1m_1}\left((\omega_1\varphi_1(\stackrel{\rightarrow}{q})
-m_1\varphi_2(\stackrel{\rightarrow}{q})\right)\left({\gamma_0}m_1-
\omega_1-{\gamma_0}{\not\!q}_{\perp}\right)\gamma_5\,.
\end{equation}
From the reduced BS, Eqs.(\ref{eqpp}, \ref{eqmm}) with a QCD
inspired instantaneous kernel (relating to $V_s$ and $V_v$) for
the $\eta$ in the equations and by taking different traces for the
$\Gamma$ matrix, finally we obtain the couple equations for the
components $\varphi_1({\stackrel{\rightarrow}{q}})$ and
$\varphi_2({\stackrel{\rightarrow}{q}})$:
\begin{eqnarray}\label{compeq1}
&(M-2\omega_1)\omega_1\left[\omega_1\varphi_1({\stackrel{\rightarrow}{q}})+
m_1\varphi_2({\stackrel{\rightarrow}{q}})\right]\nonumber \\
&=-\int\frac{d{\stackrel{\rightarrow}{k}}}{(2\pi)^3}
\left[\left(m_1^2(V_s+2V_v)-{\stackrel{\rightarrow}{q}}\cdot
{\stackrel{\rightarrow}{k}}V_s\right)\varphi_1({\stackrel{\rightarrow}{k}})+
\omega_1m_1(V_s-4V_v)\varphi_2({\stackrel{\rightarrow}{k}})\right]\,,
\end{eqnarray}
\begin{eqnarray}\label{compeq2}
&(M+2\omega_1)\omega_1\left[\omega_1\varphi_1({\stackrel{\rightarrow}{q}})-
m_1\varphi_2({\stackrel{\rightarrow}{q}})\right]\nonumber \\
&=-\int\frac{d{\stackrel{\rightarrow}{k}}}{(2\pi)^3}
\left[\left(m_1^2(V_s+2V_v)-{\stackrel{\rightarrow}{q}}\cdot
{\stackrel{\rightarrow}{k}}V_s\right)\varphi_1({\stackrel{\rightarrow}{k}})-
\omega_1m_1(V_s-4V_v)\varphi_2({\stackrel{\rightarrow}{k}})\right]\,,
\end{eqnarray}
i.e. coupled equations for the eigenvalue $M$ and eigenfunctions,
the wave functions $\varphi_1({\stackrel{\rightarrow}{q}})$ and
$\varphi_2({\stackrel{\rightarrow}{q}})$. Note that we will derive
and solve the equations with kernels $V_s$ (corresponding to
scaler confinement) and $V_v$ (corresponding to one-gluon
exchange) precisely elsewhere\cite{changwang}, but here we only
show the fact that when all of the equations Eqs.(\ref{eqpp},
\ref{eqmm}, \ref{eqpm}) are taken into account, consist coupled
equations can be obtained finally without approximation.

By contrary, for Salpeter equation only the positive energy
component equation Eq.(\ref{eqpp}) is kept, but further assumption
on the spin structure (in four components) of the wave function
$\varphi_{^{1}S_0}(\stackrel{\rightarrow}{q})$ is made\cite{salp},
that is different from here, as shown, the spin structure is
determined by the equations Eqs.(\ref{eqpp}, \ref{eqmm},
\ref{eqpm}) simultaneously.

In fact, the above coupled equations are written in the reference
frame of the bound state itself. Whereas, to give the
`instantaneous' formulation for the transitions between the bound
states, we also need to have the precise formulation for the BS
equation (with `instantaneous' kernel) in a reference frame of
another bound state. It is similar to do as the above, so the
precise formulation is put into Appendix A.

\subsection{The Reduced Formulation for Radiative Transitions between
the Bound States with Instantaneous Kernel}

The transition matrix element for a process of a bound state
emitting a photon can be written by the BS wave function as:
\begin{equation}
<P'|J_\mu|P>=<P'|J_\mu|P>_{1}+<P'|J_\mu|P>_{2}\,,
\end{equation}
where
\begin{eqnarray}
<P'|J_\mu|P>_{1}&=&\int\frac{d^4qd^4q'}{(2\pi)^{4}}Tr\left\{
\bar{\chi}_{P'}Q_{1} \gamma_{\mu} {\chi}_{P} S^{-1}_{2}(-p_2)
\delta^4(p_2-p'_2)\right\} \,, \nonumber \\
<P'|J_\mu|P>_{2}&=&\int\frac{ d^4qd^4q'}{(2\pi)^{4}}Tr\left\{
\bar{\chi}_{P'}S^{-1}_{1}(p_1){\chi}_{P}Q_{2} \gamma_{\mu}
\delta^4(p_1-p'_1) \right\}\,,
\end{eqnarray}
here we have used subscript $`1'$ to denote the photon emitted
only by the quark and $`2'$ to denote the photon emitted by the
anti-quark. In fact, when we consider the gauge invariance of the
process, these two parts are independent and each one can be
treated independently. So below when we show how to keep gauge
invariance, we pay more efforts on the first one, and just to say
similarly on the second one for shortening.

We have made instantaneous approach to the BS wave function and BS
equation, furthermore in this subsection we do similar `job' for
the transition matrix elements.

As for the matrix element
\begin{eqnarray}
&\displaystyle<P'|J_\mu|P>_{1}=Q_{1}\int\frac{
d^4qd^4q'}{(2\pi)^{4}}Tr\left\{
S'_{2}(-p'_2)\bar{\eta'}_{P'}S'_{1}(p'_1) \gamma_{\mu}
S_{1}(p_1){\eta}_{P}S_{2}(-p_2) S^{-1}_{2}(-p_2)\delta^4(p_2-p'_2)
\right\} \nonumber \\
&\displaystyle=Q_{1}\int\frac{ d^4qd^4q'}{(2\pi)^{4}}Tr\left\{
\bar{\eta'}_{P'}S'_{1}(p'_1) \gamma_{\mu}
S_{1}(p_1){\eta}_{P}S_{2} (-p_2)\delta^4(p_2-p'_2) \right\}
\nonumber \,,
\end{eqnarray}
here the integral element $d^4q=d^3q_{_P{\perp}}dq_P$ and the
$\delta$-function can be re-written by the relation:
\begin{eqnarray}
&\displaystyle\delta^4(p_2-p'_2)=
-\delta^3({p_2}_{P\perp}-{p'_2}_{P\perp})
\times\frac{1}{2\pi
i}\Bigl(\frac{1}{\alpha_{2}P_{P}-q_{P}-
\alpha'_{2}{P'}_{P}+q'_{P}+i\epsilon}
\nonumber \\
&\displaystyle-\frac{1}{\alpha_{2}P_{P}-q_{P}-
\alpha'_{2}{P'}_{P}+ q'_{P}-i\epsilon}\Bigr),
\end{eqnarray}
then with Eqs.(\ref{eqpropo}, \ref{eqpropo1}), the matrix element
turns into
$$
<P'|J_\mu|P>_{1}=-Q_{1}\int\frac{d^4qd^4q'}{(2\pi)^{4}}Tr\Bigl\{
\bar{\eta'}_{P'}S'_{1}(p'_1) \gamma_{\mu}
S_{1}(p_1){\eta}_{P}S_{2}
(-p_2)\delta^3({p_2}_{P\perp}-{p'_2}_{P\perp})$$$$
\times\frac{1}{2\pi
i}(\frac{1}{\alpha_{2}P_{P}-q_{P}-\alpha'_{2}{P'}_{P}
+q'_{P}+i\epsilon}-\frac{1}{\alpha_{2}P_{P}-q_{P}-
\alpha'_{2}{P'}_{P}+q'_{P}-i\epsilon})
\Bigr\}
$$$$ =-Q_{1}\int\frac{ d^4qd^4q'}{(2\pi)^{4}}Tr\left\{
\bar{\eta'}_{P'} S'_{1}(p'_1) \gamma_{\mu}
\left(\frac{\Lambda^{+}_{1}}{q_{P}+\alpha_{1}M-\omega_{1}+i\epsilon}+
\frac{\Lambda^{-}_{1}}{q_{P}+\alpha_{1}M+\omega_{1}-i\epsilon}\right)
\right.$$
$$\left.\times {\eta}_{P}
\left(\frac{\Lambda^{+}_{2}}{-q_{P}+\alpha_{2}M-\omega_{2}+i\epsilon}+
\frac{\Lambda^{-}_{2}}{-q_{P}+\alpha_{2}M+\omega_{2}-i\epsilon}\right)
\delta^3({p_2}_{P\perp}-{p'_2}_{P\perp})\right.$$$$\left.
\times\frac{1}{2\pi {i}}
\left(\frac{1}{\alpha_{2}M-q_{P}-\alpha'_{2}E'+q'_{P}+i\epsilon}
-\frac{1}{\alpha_{2}M-q_{P}-\alpha'_{2}E'+q'_{P}-i\epsilon}\right)
\right\}\,.
$$
To integrate over $q_P$ and in terms of Eqs.(\ref{eqpp},
\ref{eqmm}) for the initial bound state, the above matrix element
becomes the follows
$$<P'|J_\mu|P>_{1}=-Q_{1}\int\frac{ d^3q_{P\perp} d^4q'}{(2\pi)^{4}}
\delta^3({p_2}_{P\perp}-{p'_2}_{P\perp})Tr\left\{ \bar{\eta'}_{P'}
S'_{1}(p'_1) \gamma_{\mu}
\left(-\frac{\varphi^{++}_{P}(q_{P\perp})}{M-
\alpha_{2}E'-\omega_{1}+q'_P+i\epsilon}
\right.\right.$$$$\left.\left.+
\frac{\varphi^{++}_{P}(q_{P\perp})}{-\alpha_{2}E'+\omega_{2}+q'_P-i\epsilon}
-\frac{\Lambda^{+}_{1}{\eta}_{P}\Lambda^{-}_{2}}{(M-\omega_{1}+q'_{P}-\alpha_{2}E'
+i\epsilon)(\alpha_{2}E'+\omega_{2}-q'_P-i\epsilon)}
\right.\right.$$$$\left.\left.
-\frac{\Lambda^{-}_{1}{\eta}_{P}\Lambda^{+}_{2}}{(M+\omega_{1}+q'_{P}-\alpha_{2}E'
-i\epsilon)(\alpha_{2}E'-\omega_{2}-q'_P+i\epsilon)}-
\frac{\varphi^{--}_{P}(q_{P\perp})}{-\alpha_{2}E'-\omega_{2}+q'_P+i\epsilon}
\right.\right.$$$$\displaystyle\left.\left.+
\frac{\varphi^{--}_{P}(q_{P\perp})}{M-
\alpha_{2}E'+\omega_{1}+q'_P-i\epsilon}\right) \right\}\,.
$$
To apply the formulae in appendix A and to carry out the the
cantor integration over $q'_P$, we have
$$<P'|J_\mu|P>_{1}=Q_{1}\int\frac{ d^3q_{P\perp}
d^3q'_{P\perp}}{(2\pi)^{3}}
\delta^3({p_2}_{P\perp}-{p'_2}_{P\perp})Tr\left\{
\left(\frac{\bar{\eta'}_{P'} \Lambda'^{+}_{1{P'}}(q'_{P\perp})
\gamma_{\mu}\varphi^{++}_{P}(q_{P\perp})}{E'-\omega'_{1}-\omega_{2}}
\right.\right.$$$$\left.\left.-\frac{\bar{\eta'}_{P'}
\Lambda'^{-}_{1{P'}}(q'_{P\perp})
\gamma_{\mu}\varphi^{++}_{P}(q_{P\perp})}{M-\omega_{1}-\omega'_{1}-E'}
-\frac{\bar{\eta'}_{P'}
\Lambda'^{-}_{1{P'}}(q'_{P\perp})\gamma_{\mu}
\Lambda^{+}_{1}{\eta}_{P}\Lambda^{-}_{2}}
{(M-\omega_{1}-\omega'-E' )(E'+\omega'_{1}+\omega_{2})}
\right.\right.$$$$\left.\left. +\frac{\bar{\eta'}_{P'}
\Lambda'^{+}_{1{P'}}(q'_{P\perp})\gamma_{\mu}
\Lambda^{-}_{1}{\eta}_{P}\Lambda^{+}_{2}}
{(M+\omega_{1}+\omega'-E' )(E'-\omega'_{1}-\omega_{2})}
-\frac{\bar{\eta'}_{P'} \Lambda'^{+}_{1{P'}}(q'_{P\perp})
\gamma_{\mu}\varphi^{--}_{P}(q_{P\perp})}{M+\omega_{1}+\omega'_{1}-E'}
\right.\right.$$$$\left.\left. +\frac{\bar{\eta'}_{P'}
\Lambda'^{-}_{1{P'}}(q'_{P\perp})
\gamma_{\mu}\varphi^{--}_{P}(q_{P\perp})}{E'+\omega'_{1}+\omega_{2}}
\right) \right\}\,.
$$
Here we have $\delta^3({p_2}_{P\perp}-{p'_2}_{P\perp})$,
$\omega'_2=\omega_2$,
$\Lambda'^{\pm}_{2{P'}}(q'_{P\perp})=\Lambda^{\pm}_{2{P}}(q_{P\perp})$,
the identity $1\equiv \frac{\not\!{P}}{M}\frac{\not\!{P}}{M}
=\frac{\not\!{P}}{M}\left(\Lambda^{{\prime}
+}_{2}+\Lambda^{{\prime}-}_{2}\right)$ and the equations
Eqs.(\ref{eqpp}, \ref{eqmm}) for the final bound state, to apply
all of them to the matrix element, we finally obtain:
\begin{eqnarray}\label{elem_1}
&\displaystyle<P'|J_\mu|P>_{1}=Q_{1}\int\frac{ d^3q_{P\perp}
d^3q'_{P\perp}}{(2\pi)^{3}}
\delta^3({p_2}_{P\perp}-{p'_2}_{P\perp})Tr\left\{
\frac{\not\!{P}}{M}\left[
\bar{\varphi'}^{++}_{P'}\gamma_{\mu}{\varphi}^{++}_{P} +\frac{
\bar{\varphi'}^{++}_{P'}\gamma_{\mu}\Lambda^{-}_{1}
{\eta}_{P}\Lambda^{+}_{2} }{M+\omega_{1}+\omega'_{1}-E'}
\right.\right.\nonumber \\
&\displaystyle\left.\left.
-\frac{\Lambda'^{-}_{2{P'}}(q'_{P\perp})\bar{\eta'}_{P'}
\Lambda'^{+}_{1{P'}}(q'_{P\perp})
\gamma_{\mu}\varphi^{--}_{P}(q_{P\perp})}{M+\omega_{1}+\omega'_{1}-E'}
-\frac{\Lambda'^{+}_{2{P'}}(q'_{P\perp})\bar{\eta'}_{P'}
\Lambda'^{-}_{1{P'}}(q'_{P\perp})
\gamma_{\mu}\varphi^{++}_{P}(q_{P\perp})}{M-\omega_{1}-\omega'_{1}-E'}
\right.\right.\nonumber \\
&\displaystyle\left.\left. +\frac{
\bar{\varphi'}^{--}_{P'}\gamma_{\mu}\Lambda^{+}_{1}
{\eta}_{P}\Lambda^{-}_{2} }{M-\omega_{1}-\omega'_{1}-E'}-
\bar{\varphi'}^{--}_{P'}\gamma_{\mu}{\varphi}^{--}_{P}
 \right]
\right\}\,.
\end{eqnarray} If we define:
\begin{eqnarray}
\label{eq-+}{\psi}_{1P}^{-+}\equiv \frac{
\Lambda^{-}_{1}(q_{P\perp}) {\eta}_{P}\Lambda^{+}_{2}(q_{P\perp})
}{M+\omega_{1}+\omega'_{1}-E'}\,,
\end{eqnarray}
\begin{eqnarray}
\label{eq'-+} \bar{\psi}_{1P'}^{\prime-+}\equiv
\frac{\Lambda'^{-}_{2{P'}}(q'_{P\perp})
\bar{\eta}'_{P'}\Lambda'^{+}_{1{P'}}(q'_{P\perp})
}{M+\omega_{1}+\omega'_{1}-E'}\,,
\end{eqnarray}
\begin{eqnarray}
\label{eq'+-} \bar{\psi}_{1P'}^{\prime+-}\equiv
\frac{\Lambda'^{+}_{2{P'}}(q'_{P\perp})
\bar{\eta'}_{P'}\Lambda'^{-}_{1{P'}}(q'_{P\perp})
}{M-\omega_{1}-\omega'_{1}-E'}\,,
\end{eqnarray}
\begin{eqnarray}
\label{eq+-}{\psi}_{1P}^{+-}\equiv \frac{
\Lambda^{+}_{1}(q_{P\perp}) {\eta}_{P}\Lambda^{-}_{2}(q_{P\perp})
}{M-\omega_{1}-\omega'_{1}-E'}\,,
\end{eqnarray}
the amplitude becomes
\begin{eqnarray}\label{eqthree}
&\displaystyle<P'|J_\mu|P>_{1}=Q_{1}\int\frac{ d^3q_{P\perp}
d^3q'_{P\perp}}{(2\pi)^{3}}
\delta^3(-{q}_{P\perp}-\alpha_2{P'}_{P\perp}+{q'}_{P\perp})
Tr\left\{ \frac{\not\!{P}}{M}\left[
\bar{\varphi'}^{++}_{P'}({q'}_{P\perp})
\gamma_{\mu}{\varphi}^{++}_{P}({q}_{P\perp})
\right.\right.\nonumber\\
&\displaystyle\left.\left.+\bar{\varphi'}^{++}_{P'}({q'}_{P\perp})
\gamma_{\mu}{\psi}^{-+}_{P}({q}_{P\perp}) -
\bar{\psi'}^{-+}_{P'}({q'}_{P\perp})
\gamma_{\mu}{\varphi}^{--}_{P}({q}_{P\perp})
-\bar{\psi'}^{+-}_{P'}({q'}_{P\perp})
\gamma_{\mu}{\varphi}^{++}_{P}({q}_{P\perp})
\right.\right.\nonumber \\
\displaystyle&\left.\left.
+\bar{\varphi'}^{--}_{P'}({q'}_{P\perp})
\gamma_{\mu}{\psi}^{+-}_{P}({q}_{P\perp})
-\bar{\varphi'}^{--}_{P'}({q'}_{P\perp})
\gamma_{\mu}{\varphi}^{--}_{P}({q}_{P\perp})
 \right]
\right\}\,.
\end{eqnarray}
To integrate over three dimensional $q'_{P\perp}$ further, then we
obtain the instantaneous transition matrix element
$<P'|J_\mu|P>_{1}$ in three dimensional formulation. To do the
same derivation for $<P'|J_\mu|P>_{2}$, the whole instantaneous
transition matrix element becomes
\begin{eqnarray}
&<P'|J_\mu|P>=<P'|J_\mu|P>_{1}+<P'|J_\mu|P>_{2}\nonumber \\
&\displaystyle= \int\frac{ d^3q_{P\perp}}{(2\pi)^{3}}
Tr\left\{Q_{1} \frac{\not\!{P}}{M}\left[
\bar{\varphi'}^{++}_{P'}({q}_{P\perp}+\alpha_2{P'}_{P\perp})
\gamma_{\mu}{\varphi}^{++}_{P}({q}_{P\perp})+
\bar{\varphi'}^{++}_{P'}({q}_{P\perp}+\alpha_2{P'}_{P\perp})
\gamma_{\mu}{\psi}^{-+}_{1P}({q}_{P\perp}) \right.\right.
\nonumber \\
&-\left.\left.
\bar{\psi'}^{-+}_{1P'}({q}_{P\perp}+\alpha_2{P'}_{P\perp})
\gamma_{\mu}{\varphi}^{--}_{P}({q}_{P\perp})
-\bar{\psi'}^{+-}_{1P'}({q}_{P\perp}+\alpha_2{P'}_{P\perp})
\gamma_{\mu}{\varphi}^{++}_{P}({q}_{P\perp})
\right.\right.\nonumber \\
&+\left.\left.
\bar{\varphi'}^{--}_{P'}({q}_{P\perp}+\alpha_2{P'}_{P\perp})
\gamma_{\mu}{\psi}^{+-}_{1P}({q}_{P\perp})
-\bar{\varphi'}^{--}_{P'}({q}_{P\perp}+\alpha_2{P'}_{P\perp})
\gamma_{\mu}{\varphi}^{--}_{P}({q}_{P\perp}) \right]
 \right.\nonumber \\
&+\left. Q_{2} \left[
\bar{\varphi'}^{++}_{P'}({q}_{P\perp}-\alpha_1{P'}_{P\perp})
\frac{\not\!{P}}{M}{\varphi}^{++}_{P}({q}_{P\perp})
+\bar{\varphi'}^{++}_{P'}({q}_{P\perp}-\alpha_1{P'}_{P\perp})
\frac{\not\!{P}}{M}{\psi}^{+-}_{2P}({q}_{P\perp})
\right.\right.\nonumber \\
&-\left.\left.
\bar{\psi'}^{+-}_{2P'}({q}_{P\perp}-\alpha_1{P'}_{P\perp})
\frac{\not\!{P}}{M}{\varphi}^{--}_{P}({q}_{P\perp})
-\bar{\psi'}^{-+}_{2P'}({q}_{P\perp}-\alpha_1{P'}_{P\perp})
\frac{\not\!{P}}{M}{\varphi}^{++}_{P}({q}_{P\perp}) \right.\right.
\nonumber \\
&+\left.\left.
\bar{\varphi'}^{--}_{P'}({q}_{P\perp}-\alpha_1{P'}_{P\perp})
\frac{\not\!{P}}{M}{\psi}^{-+}_{2P}({q}_{P\perp})-
\bar{\varphi'}^{--}_{P'}({q}_{P\perp}-\alpha_1{P'}_{P\perp})
\frac{\not\!{P}}{M}{\varphi}^{--}_{P}({q}_{P\perp})
\right]\gamma_{\mu} \right\}\,. \label{ins-ele}
\end{eqnarray}
Here similar to the Eqs.(\ref{eq-+}-\ref{eq+-}), we have defined
$\psi_{2P}^{\pm \mp}$ and ${\bar\psi}_{2P'}^{\prime\pm \mp}$ as
follows:
\begin{eqnarray}
\label{eq2+-} {\psi}_{2P}^{+-}=\frac{ \Lambda^{+}_{1}(q_{P\perp})
{\eta}_{P}\Lambda^{-}_{2}(q_{P\perp})
}{M+\omega_{2}+\omega'_{2}-E'}\,,
\end{eqnarray}
\begin{eqnarray}
\label{eq'2+-}
\bar{\psi}_{2P'}^{\prime+-}=\frac{\Lambda'^{+}_{2{P'}}(q'_{P\perp})
\bar{\eta}'_{P'}\Lambda'^{-}_{1{P'}}(q'_{P\perp})
}{M+\omega_{2}+\omega'_{2}-E'}\,,
\end{eqnarray}
\begin{eqnarray}
\label{eq'2-+}
\bar{\psi}_{2P'}^{\prime-+}=\frac{\Lambda'^{-}_{2{P'}}(q'_{P\perp})
\bar{\eta'}_{P'}\Lambda'^{+}_{1{P'}}(q'_{P\perp})
}{M-\omega_{2}-\omega'_{2}-E'}\,,
\end{eqnarray}
\begin{eqnarray}
\label{eq2-+}{\psi}_{2P}^{-+}=\frac{ \Lambda^{-}_{1}(q_{P\perp})
{\eta}_{P}\Lambda^{+}_{2}(q_{P\perp})
}{M-\omega_{2}-\omega'_{2}-E'}\,.
\end{eqnarray}

If one takes the approximation
${\varphi}^{--}_{P}({q}_{P\perp})\simeq 0$, the matrix element
becomes
\begin{eqnarray}
&<P'|J_\mu|P>=<P'|J_\mu|P>_{1}+<P'|J_\mu|P>_{2}\nonumber \\
&\displaystyle= \int\frac{ d^3q_{P\perp}}{(2\pi)^{3}}
Tr\left\{Q_{1} \frac{\not\!{P}}{M}\left[
\bar{\varphi'}^{++}_{P'}({q}_{P\perp}+\alpha_2{P'}_{P\perp})
\gamma_{\mu}{\varphi}^{++}_{P}({q}_{P\perp})
\right.\right.\nonumber \\
&\displaystyle+\left.\left.
\bar{\varphi'}^{++}_{P'}({q}_{P\perp}+\alpha_2{P'}_{P\perp})
\gamma_{\mu}{\psi}^{-+}_{1P}({q}_{P\perp})
-\bar{\psi'}^{+-}_{1P'}({q}_{P\perp}+\alpha_2{P'}_{P\perp})
\gamma_{\mu}{\varphi}^{++}_{P}({q}_{P\perp})
 \right]
 \right.\nonumber \\
&\displaystyle+ \left. Q_{2} \left[
\bar{\varphi'}^{++}_{P'}({q}_{P\perp}-\alpha_1{P'}_{P\perp})
\frac{\not\!{P}}{M}{\varphi}^{++}_{P}({q}_{P\perp})
+\bar{\varphi'}^{++}_{P'}({q}_{P\perp}-\alpha_1{P'}_{P\perp})
\frac{\not\!{P}}{M}{\psi}^{+-}_{2P}({q}_{P\perp})
\right.\right.\nonumber \\
&\displaystyle-\left.\left.
\bar{\psi'}^{-+}_{2P'}({q}_{P\perp}-\alpha_1{P'}_{P\perp})
\frac{\not\!{P}}{M}{\varphi}^{++}_{P}({q}_{P\perp})
\right]\gamma_{\mu} \right\}\,. \label{ins-eles0}
\end{eqnarray}

Note that even when the approximation
${\varphi}^{--}_{P}({q}_{P\perp})\simeq 0$ is made, from
Eq.(\ref{ins-eles0}) one may see that it is still not correct to
calculate the matrix element by means of
\begin{eqnarray}
&<P'|J_\mu|P>=<P'|J_\mu|P>_{1}+<P'|J_\mu|P>_{2}\nonumber \\
&\displaystyle= \int\frac{ d^3q_{P\perp}}{(2\pi)^{3}}
Tr\left\{Q_{1} \frac{\not\!{P}}{M}\left[
\bar{\varphi'}^{++}_{P'}({q}_{P\perp}+\alpha_2{P'}_{P\perp})
\gamma_{\mu}{\varphi}^{++}_{P}({q}_{P\perp}) \right]
\right.\nonumber \\
&\displaystyle+ \left. Q_{2} \left[
\bar{\varphi'}^{++}_{P'}({q}_{P\perp}-\alpha_1{P'}_{P\perp})
\frac{\not\!{P}}{M}{\varphi}^{++}_{P}({q}_{P\perp})
\right]\gamma_{\mu} \right\}\,. \label{ins-eles}
\end{eqnarray}

\subsection{The Gauge Invariance of Instantaneous Transition Matrix
Element}

In Sec.II, we showed that the transition matrix element is gauge
invariant no matter the instantaneous approach is applied to,
i.e.,
\begin{eqnarray}
\displaystyle-iQ^{\mu}<P'|J_\mu|P>_1=\frac{1}{(2\pi)^4}\int
d^4q'\int d^4qTr\left\{\bar{\chi}_{P'}(q')
Q_1\left([S_{f}^{(1)}(\alpha_1P+q)]^{-1} \right.\right.\nonumber\\
\left.\left.-[S_{f}^{(1)}(\alpha'_1P'+q')]^{-1}\right)
{\chi}_{P}(q)[S^{(2)}_{f}(-\alpha'_2P'+q')]^{-1}
\delta^{4}(q-q'-\alpha_2Q)\right\}=0\,.
\label{eqgau}
\end{eqnarray}
In this subsection, we show that the instantaneous transition
matrix element Eq.(\ref{ins-ele}) is also gauge invariant, in
terms of the gauge invariance in four dimensional formulation
Eq.(\ref{eqgau} and the instantaneous equations
Eqs.(\ref{eqpp},\ref{eqmm},\ref{eqpm}).

First of all, let us examine how the equation Eq.(\ref{eqgau})
turns to a three dimensional one when the kernel of the considered
BS equation is instantaneous. To substitute the BS wave functions
$\chi_{P}(q)$ and $\bar{\chi}_{P'}(q')$ with BS equation
Eq.(\ref{eqBSin}), and similar to the above subsections, to deal
with $\delta$ functions and then to integrate out $q_P$ and $q'_P$
in the equation by contour integral, one immediately obtain
\begin{eqnarray}
\label{eq37}
&\displaystyle-iQ^{\mu}<P'|J_\mu|P>_1=\int\frac{
d^4qd^4q'}{(2\pi)^{4}}Tr\left\{ S_f^{'(2)}(-p'_2)\bar{\eta}_{P'}
S_f^{'(1)}(p'_1) {\eta}_{P}-\bar{\eta}_{P'}
S_f^{(1)}(p_1){\eta}_{P}
S_f^{(2)}(-p_2)\right\} \nonumber \\
&\displaystyle\times \delta^3({p_2}_{P\perp}-{p'_2}_{P\perp})
\delta(\alpha_2M-q_P-\alpha_2{P'}_P+q'_P)\nonumber \\
&\displaystyle=-{2\pi i}\int\frac{
d^3q_{\perp}d^3q'_{\perp}\delta^3({p_2}_{\perp}
-{p'_2}_{\perp})}{(2\pi)^{4}} Tr\left\{ \bar{\varphi}'^{++}_{P'}
{\eta}_{P} -\bar{\varphi}'^{--}_{P'} {\eta}_{P}-\bar{\eta}_{P'}
{\varphi}^{++}_{P_i} +\bar{\eta}_{P'} {\varphi}^{--}_{P}
\right\}\,.
\end{eqnarray}

Note here that with Eqs.(\ref{eqpp}, \ref{eqmm}), we have the part
in $Tr$ of Eq.(\ref{eq37}):
\begin{eqnarray}
&\displaystyle\left\{ \bar{\varphi}'^{++}_{P'} {\eta}_{P}
-\bar{\varphi}'^{--}_{P'} {\eta}_{P}-\bar{\eta}_{P'}
{\varphi}^{++}_{P_i} +\bar{\eta}_{P'} {\varphi}^{--}_{P}
\right\}\nonumber \\
&\displaystyle=\left\{\frac{\Lambda'^{+}_{2{P'}}(q'_{P\perp})
\bar{\eta'}_{P'}\Lambda'^{+}_{1{P'}}(q'_{P\perp})
}{E'-\omega_{2}-\omega'_{1}}{\eta}_{P}
+\frac{\Lambda'^{-}_{2{P'}}(q'_{P\perp})
\bar{\eta'}_{P'}\Lambda'^{-}_{1{P'}}(q'_{P\perp})}
{E'+\omega_{2}+\omega'_{1}} {\eta}_{P}\right.\nonumber \\
&\displaystyle\left.-\bar{\eta'}_{P'}
\frac{\Lambda^{+}_{1{P}}(q_{P\perp})
{\eta}_{P}\Lambda^{+}_{2{P}}(q_{P\perp})}
{M-\omega_{2}-\omega'_{1}}
-\bar{\eta'}_{P'}\frac{\Lambda^{-}_{1{P}}(q_{P\perp})
{\eta}_{P}\Lambda^{+}_{2{P'}}(q_{P\perp})}
{M+\omega_{2}+\omega'_{1}} \right\}\,. \label{xyz}
\end{eqnarray}
From Eq.(\ref{xyz}), one cannot be sure the difference between the
second and the fourth terms will be smaller than the difference
between the first and the third terms, although each of the first
and the third terms itself is big and each of the second and the
fourth terms is small. Practically, we have taken several examples
for heavy quarkonium, and indeed we have found when
${\varphi}'^{--}_{P'}$ and ${\varphi}^{--}_{P}$ in Eq.(\ref{eq37})
are neglected, then we cannot obtain satisfied results with gauge
invariance. Therefore, when one would like to have the gauge
invariance for the transition matrix elements, one cannot ignore
any of the `negative energy' components in Eq.(\ref{ins-ele}).

By means of the relations of projection operators Eq.(\ref{eq24}),
the above equation Eq.(\ref{eqgau}) becomes:
$$-iQ^{\mu}<P'|J_\mu|P>_1=-{2\pi
i}\int\frac{
d^3q_{\perp}d^3q'_{\perp}\delta^3({p_2}_{\perp}-{p'_2}_{\perp})}{(2\pi)^{4}}
Tr\left\{ \bar{\varphi}'^{++}_{P'}
\frac{\not\!{P}}{M}(\Lambda^{+}_{1}+\Lambda^{-}_{1}){\eta}_{P}
(\Lambda^{+}_{2}+\Lambda^{-}_{2})\frac{\not\!{P}}{M}\right.$$
$$\left.
-\bar{\varphi}'^{--}_{P'}
\frac{\not\!{P}}{M}(\Lambda^{+}_{1}+\Lambda^{-}_{1}){\eta}_{P}
(\Lambda^{+}_{2}+\Lambda^{-}_{2})\frac{\not\!{P}}{M}-
\frac{\not\!{P}}{M}(\Lambda'^{+}_{2}+\Lambda'^{-}_{2})\bar{\eta}_{P'}
(\Lambda'^{+}_{1}+\Lambda'^{-}_{1})\frac{\not\!{P}}{M}
{\varphi}^{++}_{P} \right.$$
$$\left.+\frac{\not\!{P}}{M}(\Lambda'^{+}_{2}+
\Lambda'^{-}_{2})\bar{\eta}_{P'}
(\Lambda'^{+}_{1}+\Lambda'^{-}_{1})\frac{\not\!{P}}{M}
{\varphi}^{--}_{P} \right\}=0\,.$$ Furthermore, with
Eqs.(\ref{eqpp},\ref{eqmm},\ref{eq-+}-\ref{eq+-}), we finally
have:
\begin{eqnarray}
&&-iQ^{\mu}<P'|J_\mu|P>_1=-\frac{i}{(2\pi)^3}\int
d^3q_{\perp}Tr\left\{\left[ \bar{\varphi'}^{++}_{P'}
\frac{\not\!{P}}{M}\varphi^{++}(M-
\omega_1-\omega_2-E'+\omega'_{1}+\omega'_2)
\right.\right.\nonumber \\
&&\left.\left.+\bar{\varphi'}^{++}_{P'}
\frac{\not\!{P}}{M}\psi^{-+}(M+ \omega_1+\omega'_{1}-E')-
\bar{\varphi'}^{--}_{P'} \frac{\not\!{P}}{M}\psi^{+-}(M-
\omega_1-\omega'_{1}-E') \right.\right. \nonumber \\
&&\left.\left.-\bar{\psi'}^{+-}_{P'}
\frac{\not\!{P}}{M}\varphi^{++}(M- \omega_1-\omega'_{1}-E')+
\bar{\psi'}^{-+}_{P'} \frac{\not\!{P}}{M}\varphi^{--}(M+
\omega_1+\omega'_{1}-E')\right.\right. \nonumber \\
&&\left.\left.- \bar{\varphi'}^{--}_{P'}
\frac{\not\!{P}}{M}\varphi^{--}(M+
\omega_1+\omega_2-E'-\omega'_{1}-
\omega'_2)\right]\frac{\not\!{P}}{M}\right\}=0\,. \label{eq38}
\end{eqnarray}
Namely, Eq.(\ref{eq38}) is the representation of electromagnetic
gauge invariance (the current conservation, $-iQ^{\mu}
<P'|J_\mu|P>_1=0$) for the instantaneous (three-dimensional)
formulation.

Now let us show a given instantaneous transition matrix element,
e.g., Eq.(\ref{eqthree}) indeed is gauge invariant.

If we multiply $-iQ^{\mu}$ to Eq.(\ref{eqthree}) and apply :
\begin{eqnarray}
&\displaystyle-iQ^{\mu}Q_{1}\int\frac{ d^3q_{P\perp}}{(2\pi)^{3}}
Tr\left\{ \frac{\not\!{P}}{M}\left[ \bar{\varphi'}^{++}_{P'}
\gamma_{\mu}{\varphi}^{++}_{P} +\bar{\varphi'}^{++}_{P'}
\gamma_{\mu}{\psi}^{-+}_{P} + \bar{\psi'}^{-+}_{P'}
\gamma_{\mu}{\varphi}^{--}_{P}\right.\right. \nonumber \\
&\displaystyle\left.\left. -\bar{\psi'}^{+-}_{P'}
\gamma_{\mu}{\varphi}^{++}_{P}
 -\bar{\varphi'}^{--}_{P'}
\gamma_{\mu}{\psi}^{+-}_{P} -\bar{\varphi'}^{--}_{P'}
\gamma_{\mu}{\varphi}^{--}_{P} \right] \right\}\nonumber
\end{eqnarray}
we have
\begin{eqnarray}
&\displaystyle-iQ_{1}\int\frac{ d^3q_{P\perp}}{(2\pi)^{3}}
Tr\left\{ \frac{\not\!{P}}{M}\left[ \bar{\varphi'}^{++}_{P'}
(\not\!{P}-\not\!{P'}){\varphi}^{++}_{P} +\bar{\varphi'}^{++}_{P'}
(\not\!{P}-\not\!{P'}){\psi}^{-+}_{P} + \bar{\psi'}^{-+}_{P'}
(\not\!{P}-\not\!{P'}){\varphi}^{--}_{P}\right.\right. \nonumber \\
&\displaystyle\left.\left. -\bar{\psi'}^{+-}_{P'}
(\not\!{P}-\not\!{P'}){\varphi}^{++}_{P}
 -\bar{\varphi'}^{--}_{P'}
(\not\!{P}-\not\!{P'}){\psi}^{+-}_{P} -\bar{\varphi'}^{--}_{P'}
(\not\!{P}-\not\!{P'}){\varphi}^{--}_{P} \right]\,.
\right\}\label{ingauge}
\end{eqnarray}

With the properties Eqs.(\ref{appa11}, \ref{appa12}, \ref{appa13})
for the project operators, we may obtain the identities:
\begin{eqnarray}
&\not\!{P}-\not\!{P'}=\not\!{p_1}-\not\!{p'_1}=
[2\omega_1\Lambda^{+}_{1}+
\frac{\not\!{P}}{M}(\alpha_1M-\omega_1+q_P)]
-[2\omega'_1\Lambda'^{+}_{1}+
\frac{\not\!{P}}{M}(\alpha_1E'-\omega'_1+q'_P)]
\nonumber \\
&=[-2\omega_1\Lambda^{-}_{1}+
\frac{\not\!{P}}{M}(\alpha_1M+\omega_1+q_P)]
-[-2\omega'_1\Lambda'^{-}_{1}+
\frac{\not\!{P}}{M}(\alpha_1E'+\omega'_1+q'_P)]\,,
\end{eqnarray}
\begin{eqnarray}
\Lambda'^{+}_{1}(\not\!{P}-\not\!{P'})\Lambda^{+}_{1}
=\Lambda'^{+}_{1}\frac{\not\!{P}}{M}\Lambda^{+}_{1}(M-
\omega_1-\omega_2-E'+\omega'_{1}+\omega'_2)\,,
\end{eqnarray}

\begin{equation}\Lambda'^{+}_{1}(\not\!{P}-\not\!{P'})
\Lambda^{-}_{1}
=\Lambda'^{+}_{1}\frac{\not\!{P}}{M}\Lambda^{-}_{1}
(M+\omega_1+\omega'_{1}-E')\,,
\end{equation}

\begin{equation}\Lambda'^{-}_{1}(\not\!{P}-
\not\!{P'})\Lambda^{+}_{1}
=\Lambda'^{-}_{1}\frac{\not\!{P}}{M}\Lambda^{+}_{1}
(M-\omega_1-\omega'_{1}-E')\,,
\end{equation}

\begin{equation}\Lambda'^{-}_{1}(\not\!{P}-\not\!{P'})\Lambda^{-}_{1}
=\Lambda'^{-}_{1}\frac{\not\!{P}}{M}\Lambda^{-}_{1}(M+
\omega_1+\omega_2-E'-\omega'_{1}-\omega'_2)\,.
\end{equation}
To apply the above identities for the project operators and with
the definitions Eqs.(\ref{eq+-}, \ref{eq'+-}, \ref{eq-+},
\ref{eq'-+}), one can turn the equation Eq.(\ref{ingauge}) to the
equation Eq.(\ref{eq38}), the terms proportional to $Q_1$
accordingly. For the terms proportional to $Q_2$ it is the same.
This means the gauge invariance (current conservation) is
satisfied, and also show that it does not matter to multiply
$-iQ^{\mu}$ to Eq.(\ref{eqthree}) first and then to make the
contour integration, or to reverse the order of the two steps,
even if this two ways to do the contour integration may meet
different poles.

\section{Discussions and outlook}

As shown in the derivations, it is not so straightforward to
guarantee the gauge invariance for the instantaneous matrix
elements, so one should be full of confidence to apply them to
various processes. We derive the Eqs.(\ref{eqthree},
\ref{ins-ele}) without approximation and to start with a full
gauge conserved four-dimensional formulation on BS wave functions,
no matter the BS kernel being instantaneous or not (just carry out
the contour integrations when the BS kernel is instantaneous), so
it is understandable that the final instantaneous matrix elements
Eqs.(\ref{eqthree}, \ref{ins-ele}) are gauge invariant (the matrix
element for current is conserved). Since a physical process, no
matter it contains bound state(s) or not, must be gauge invariant,
so from the equations here one may conclude that the components
correspond to the negative energy projection in the BS wave
function should be also treated carefully, even the BS equation
with an instantaneous interaction, no matter one wishes full
relativistic effects are taken into account or not. If only the
positive energy component, although being the biggest one in wave
function, is kept, one can not have a controlled result, because
it violates gauge invariance and most of the processes are
sensitive to the gauge choice.

If in addition to the `positive energy' component equation
Eq.(\ref{eqpp}) which Salpeter kept only, the `negative energy'
component equation Eq.(\ref{eqmm}) and the `constraint equations'
Eq.(\ref{eqpm}) between the positive and negative energy ones for
the BS wave function, as emphasized here, all are applied
simultaneously, then the final equations corresponding to the
scalar components of the full BS wave function become a group of
consistent equations being coupled each other. The fact is shown
by the example Eqs.(\ref{compeq1}, \ref{compeq2}) clearly.

If keeping all the components of the BS equation, including the
`negative energy' components Eq.(\ref{eqmm}) with the `constraint'
Eq.(\ref{eqpm}), then it is indicated that accordingly the
effective theories, such as NRQED\cite{nrqed} and
NRQCD\cite{braat} etc, should be affected in certain degree. It is
because that the effective theories, such as NRQED and NRQCD, are
established based on the considerations about the symmetries,
gauge invariance etc, but the components (sectors) corresponding
to the fermion and anti-fermion field are decoupled completely at
beginning\cite{nrqed,braat}. Indeed the effective theories can be
expended in $O(v)$ (where $v=\frac{|\overrightarrow{p}|}{m}$ is
the relative velocity) accurately and the gauge invariance may be
kept precisely for each sector itself of the fermion and the
anti-fermion respectively, but they can never include any of the
special `relativistic effects' which correspond to the relations
and/or constraints between the fermion and anti-fermion sectors
i.e. only the `positive energy' component equation Eq.(\ref{eqpp})
itself can involve some of them, but cannot involve those which
come from the `negative energy' component equation Eq.(\ref{eqmm})
and the constraint equation Eq.(\ref{eqpm}).

For relativistic quantum field theories, the fermion and its
anti-particle are related definitely. Dirac spinor (four
components) relates the fermion and its anti-particle sector
properly but Pauli one cannot. NRQCD and NRQED are established at
beginning on the Pauli spinors for the heavy quark sectors with
fermion and anti-fermion sectors completely being decoupled, so
they cannot involve all of the connections from the relations and
and constraints such as Eqs (\ref{eqmm}, \ref{eqpm}) besides
Eq.(\ref{eqpp}). Note that since here it is the `specially' case
with instantaneous interaction shown, so the precise relations
and/or constraints are determined by Eqs.(\ref{eqpp}, \ref{eqmm}),
\ref{eqpm}, whereas in general cases they should be restricted not
as here with the `instantaneous interaction. If one would not
ignore the negative energy components and wish to involve the
effects in non-relativistic effective theories on corresponding
underlying ones such as the BS equations i.e. on relativistic
quantum field theories for the bound states, the effective
theories should have certain corrections to dictate the relations
and/or constraints between the sectors of the fermion and its
anti-particle. Namely even each of the sectors for the fermion and
its anti-particle itself is a non-relativistic with quite accurate
corrections in $O(v)$ and gauge invariant, but certain relations
and or constrains between the fermion and its anti-particle
sectors should be involved in correctly, especially, when the
relations and/or constraints are put on, then the gauge invariance
should be well-guaranteed, as long as one does consider the
fermion and anti-fermion to form bound states and wish to reach to
a high accuracy for transition calculations between the bound
states. As indicated by Eqs.(\ref{constr0}, \ref{0-+},
\ref{compeq1}, \ref{compeq2}) (although they are obtained when the
binding interaction is instantaneous0, this kind corrections
focused here for the effective theories vary with the quantum
numbers, such as the quantum numbers $J^{PC}$ (when the bound
states are formed by a fermion and its anti-particle exactly) of
the bound states. We will discuss this topic more precisely
elsewhere\cite{wchang2}.

We also think that the instantaneous formulations obtained in the
paper for the radiative transitions and the BS equation should
have various tests, e.g., to solve the coupling equations obtained
from the BS equations with an instantaneous kernel, and to apply
the obtained solutions (the spectrum and the corresponding wave
functions) to the radiative transitions etc. Such applications of
the formulations to positronium and charmonium as precise examples
are in progress, and preliminary numerical results have been
obtained already\cite{changwang}.

\vspace*{0.6cm} \noindent {\Large\bf Acknowledgements}

\vspace*{0.6cm} \noindent

The author (C.H. Chang) would like to thank Prof. Stephen L. Adler
for valuable discussions and to thank Prof. Xiangdong Ji for very
useful discussions. He also would like to thank Prof. Stephen L.
Adler and Institute for Advanced Study for the warmest hospitality
during his visit of the institute because the paper was completed
there. This work was supported in part by the National Natural
Science Foundation of China.


\vspace{2cm}

\appendix{

\noindent
{\Large\bf Appendices}

\section{The General Formulation for the `Instantaneous' BS
Equation}

Generally, a non-relativistic binding system can form several
(even infinity) bound states, i.e., the BS equation have several
(even infinity) solutions to correspond the spectrum of the
binding system. When possible transitions among them are
concerned, and the system is `instantaneous', we need to formulate
the `instantaneous' BS equation and the corresponding BS wave
functions not only on the solution itself but also on a different
solution from it. Namely the initial and final bound states, all
of them should be formulated in one reference frame, i.e., both of
them are either in the reference frame of the initial state $P$
(with BS wave function $\chi_P(q)$) or in that of the final state
$P'$ (with BS wave function $\chi_{P'}(q')$).

Considering the applications, here let us describe the precise
formulation of the `instantaneous' BS equation and the relevant BS
wave functions $\chi_P(q{^{\mu}})$ and $\chi_{P'}(q'{^{\mu}})$ not
only on the state $P$ but also on another one $P'$. In the text of
the paper, we have established the `instantaneous' formulation of
the BS equation `set on' the state $P$, which may be moving or
rest to the reference frame. Now we need to establish the
formulation of the BS equation for the state $P'$, but again in
the reference frame of the state $P$ (another BS equation solution
for the bound state $P'$). Thus we need to divide the momentum
$P'$ and $q'$, as done for the relative momentum $q$ for
$\chi_P(q^\mu)$, into two components: a parallel component
($P'_{_P\parallel}$, $q'_{_P\parallel}$) and an orthogonal one
($P'_{_P\perp}$, $q'_{_P\perp}$) to the momentum $P$ (the momentum
of another bound state) respectively. For convenience, we write
the divisions for $P'$ and $q'$ as well below:
$$P'{^{\mu}}=P'{^{\mu}}_{_P\parallel}+P'{^{\mu}}_{_P\perp},$$
$$q'^{\mu}=q'^{\mu}_{_P\parallel}+q'^{\mu}_{_P\perp},$$
where $P'{^{\mu}}_{_P\parallel}\equiv (P\cdot P'/M^{2})P^{\mu}$,
$P'{^{\mu}}_{_P\perp}\equiv P'{^{\mu}}-P'{^{\mu}}_{_P\parallel}$,
$q'^{\mu}_{_P\parallel}\equiv (P\cdot q'/M^{2})P^{\mu}$ and
$q'^{\mu}_{_P\perp}\equiv q'^{\mu}-q'^{\mu}_{_P\parallel}$.
Correspondingly, we have the Lorentz invariant variables:
$$P'_{P}=\frac{P\cdot P'}{M}\,,\;\;\;\;
P'_{P_T}=\sqrt{P'{^{2}}_{P}-P'{^{2}}}=\sqrt{-P'{^{2}}_{_P\perp}}\,,$$
$$q'_{P}=\frac{P\cdot q'}{M_{P}}\,,\;\;\;\;
q'_{P_T}=\sqrt{q'{^{2}}_{P}-q'{^{2}}}=\sqrt{-q'^{2}_{_P\perp}}\,.$$
The `instantaneous' interaction kernel is the same as
Eq.(\ref{eqinst}):
\begin{equation}
\label{eqinst2} V(P^{\mu},k^\mu,q'^\mu)=
V(|k^{\mu}_{P_\perp}-q'^{\mu}_{P_\perp}|)=V(P'{^{\mu}},k{^\mu},q'{^\mu})\,.
\end{equation}
Thus the relevant `instantaneous' BS wave function and equation
are
\begin{equation}
\label{eqinstw1} \varphi_{P'}(q'^{\mu}_{_P\perp})\equiv i\int
\frac{dq'_{P}}{2\pi}\chi_{P'}(q'^{\mu}_{_P\parallel},q'^{\mu}_{_P\perp})\,,
\end{equation}
and integration
\begin{equation}
\eta'(q'^{\mu}_{_P\perp})\equiv\int\frac{k^{2}_{P_T}dk_{P_T}ds}{(2\pi)^{2}}
V(|k^\mu_{_P\perp}-q'^\mu_{_P\perp}|)\varphi_{P'}(k^{\mu}_{_P\perp})\,,
\end{equation}
then the BS equation becomes
\begin{equation}
\label{eqBSin1}
\chi_{P'}(q'^\mu_{_P\parallel},q'^\mu_{_P\perp})=S_f^{(1)}(p'_1{^\mu})
\eta'(q'^\mu_{_P\perp})S_f^{(2)}(-p'_2{^\mu})\,,
\end{equation}
where $S_f^{(1)}(p'_1{^\mu})$ and $S_f^{(2)}(-p'_2{^\mu})$ are the
propagators of the quark and anti-quark respectively with the
momenta Eq.(\ref{eqqp}). The propagators $S_f^{(1)}(p'_1),
S_f^{(2)}(-p'_2)$ and relevant projection operators of the bound
state $P'$ but in the frame of $P$ can be written as follows.
\begin{equation}
iJS_f^{(i)}(Jp'{^\mu}_{i})=\frac{\Lambda'{^{+}}_{iP}(q'^\mu_{_P\perp})}{Jq'_{P}
+\alpha'_{i}P'_P-\omega'_{iP}+i\epsilon}+
\frac{\Lambda'{^{-}}_{iP}(q'^\mu_{_P\perp})}{Jq'_{P}+\alpha'_{i}P'_P+
\omega'_{iP}-i\epsilon}\,,
\end{equation}
with
\begin{eqnarray}
&&\omega'_{1P}=\sqrt{m'_{1}{^{2}}-(q'_{_P\perp}+
\alpha'_1P'_{P\perp})^{2}}\,,\;\;\;\;
\omega'_{2P}=\sqrt{m'_{2}{^{2}}-(q'_{_P\perp}-\alpha'_2P'_{P\perp})^{2}}
\,,\nonumber \\
&&\Lambda'^{\pm}_{1P}(q'_{P_\perp})= \frac{1}{2\omega'_{1P}}\left[
\frac{\not\!\!{P}}{M}\omega'_{1P}\pm
(m'_{1}+{\not\!q'}_{_P\perp}+{\alpha'_1\not\!P'}_{_P\perp})\right]\,,\nonumber\\
&&\Lambda'^{\pm}_{2P}(q'_{P_\perp})= \frac{1}{2\omega'_{2P}}\left[
\frac{\not\!\!{P}}{M}\omega'_{2P}\mp
(m'_{2}+{\not\!q'}_{_P\perp}-{\alpha'_2\not\!P'}_{_P\perp})\right]\,.
\end{eqnarray}
Here $\Lambda'^{\pm}_{iP}(q'_{P_\perp})$ satisfies the relations:
\begin{equation}
\Lambda'^{+}_{iP}(q'^\mu_{_P\perp})+\Lambda'^{-}_{iP}(q'^\mu_{_P\perp})=
\frac{\not\!\!{P}}{M}\,,\;\;\;
\Lambda'^{\pm}_{iP}(q'^\mu_{_P\perp})\frac{\not\!\!{P}}{M}
\Lambda'^{\pm}_{iP}(q'^\mu_{_P\perp})=\Lambda'^{\pm}_{iP}(q'^\mu_{_P\perp})
\,,\;\;\;
\Lambda'^{\pm}_{iP}(q'^\mu_{_P\perp})\frac{\not\!\!{P}}{M}
\Lambda'^{\mp}_{iP}(q'^\mu_{_P\perp})=0\,.
\end{equation}

Now $\varphi^{\pm\pm}_{P'}(q'^\mu_{_P\perp})$ are defined as:
\begin{equation}
\varphi^{\pm\pm}_{P'}(q'^\mu_{_P\perp})\equiv
\Lambda'^{\pm}_{1P}(q'^\mu_{_P\perp})
\frac{\not\!\!{P}}{M}\varphi_{P'}(q'^\mu_{_P\perp})
\frac{\not\!\!{P}}{M} \Lambda'^{\pm}_{2P}(q'^\mu_{_P\perp})\,,
\end{equation}
then
$$\varphi_{P'}(q'^\mu_{_P\perp})=\varphi^{++}_{P'}(q'^\mu_{_P\perp})+
\varphi^{+-}_{P'}(q'^\mu_{_P\perp})+\varphi^{-+}_{P'}(q'^\mu_{_P\perp})
+\varphi^{--}_{P'}(q'^\mu_{_P\perp})\,.$$

Integrating out $q'_{P}$ (contour integration) on both sides of
Eq.(\ref{eqBSin1}) we obtain
\begin{eqnarray}
\label{eqcont1}
\varphi_{P'}(q'^\mu_{_P\perp})=\frac{\Lambda'{^{+}}_{1P}(q'^\mu_{_P\perp})
\eta'(q'^\mu_{_P\perp})\Lambda'{^{+}}_{2P}(q'^\mu_{_P\perp})}
{(P'_{P}-\omega'_{1P}-\omega'_{2P})}-\frac{\Lambda'{^{-}}_{1P}(q'^\mu_{_P\perp})
\eta'(q'^\mu_{_P\perp})\Lambda'{^{-}}_{2P}(q'^\mu_{_P\perp})}
{(P'_{P}+\omega'_{1P}+\omega'_{2P})}\,,
\end{eqnarray}
and applying the $P$-projection operators
$\Lambda'{^\pm}_{iP}(q'^\mu_{_P\perp})$ to Eq.(\ref{eqcont1})
further properly, we obtain the coupled equations:
\begin{eqnarray}
\label{eqpp1}
(P'_P-\omega'_{1P}-\omega'_{2P})\varphi^{++}_{P'}(q'^\mu_{_P\perp})=
\Lambda'{^{+}}_{1P}(q'^\mu_{_P\perp})
\eta'(q'^\mu_{_P\perp})\Lambda'{^{+}}_{2P}(q'^\mu_{_P\perp})\,,
\end{eqnarray}
\begin{eqnarray}
\label{eqmm1}
(P'_P+\omega'_{1P}+\omega'_{2P})\varphi^{--}_{P'}(q'^\mu_{_P\perp})=
-\Lambda'{^{-}}_{1P}(q'^\mu_{_P\perp})
\eta'(q'^\mu_{_P\perp})\Lambda'{^{-}}_{2P}(q'^\mu_{_P\perp})\,,
\end{eqnarray}
\begin{equation}
\label{eqpm1}
\varphi^{+-}_{P'}(q'^\mu_{_P\perp})=\varphi^{-+}_{P'}(q'^\mu_{_P\perp})=0\,,
\end{equation}
the similarities of the equations Eqs.(\ref{eqpp}, \ref{eqmm},
\ref{eqpm}).

\section{Useful Formulae for the Project Operators}

In this appendix, we present the formulae for the project
operators which are used in the proving the gauge invariance for
the instantaneous formulation of the radiative transitions.

When the photon is emitted by the $1$-quark $Q=p_1-p_1'$, i.e. the
$2$-quark is kept unchanged (Fig.1)
$\delta^3(-{q}_{P\perp}-\alpha_2{P'}_{P\perp}+{q'}_{P\perp})$,
then we have $\omega'_2=\omega_2$ and
$\Lambda'^{\pm}_{2}=\Lambda^{\pm}_{2}$. Furthermore it is easy to
prove the identities for the project operators
$\Lambda'^{\pm}_{1P}(q_{P_\perp})$:
\begin{equation}\label{appa11}
\Lambda'^{\pm}_{1}(q'^\mu_{_P\perp})\Lambda'^{\pm}_{1}(q'^\mu_{_P\perp})=
\pm\frac{{m_1}}{\omega'_1}\Lambda'^{\pm}_{1}(q'^\mu_{_P\perp})
\,,\;\;
\Lambda^{\pm}_{1}(q^\mu_{_P\perp})\Lambda^{\pm}_{1}(q^\mu_{_P\perp})=
\pm\frac{{m_1}}{\omega_1}\Lambda^{\pm}_{1}(q^\mu_{_P\perp})\,;
\end{equation}
\begin{equation}\label{appa12}
\Lambda'^{+}_{1}(q'^\mu_{_P\perp})\Lambda'^{-}_{1}(q'^\mu_{_P\perp})=
\frac{(\not\!{q'}_{P\perp}+\alpha_1\not\!{P'}_{P\perp})}{\omega'_1}
\Lambda'^{-}_{1}(q'^\mu_{_P\perp}) \,,\;\;
\Lambda^{+}_{1}(q^\mu_{_P\perp})\Lambda^{-}_{1}(q^\mu_{_P\perp})=
\frac{\not\!{q}_{P\perp}}{\omega_1}\Lambda^{-}_{1}(q^\mu_{_P\perp})\,;
\end{equation}
\begin{equation}\label{appa13}
\Lambda'^{-}_{1}(q'^\mu_{_P\perp})\Lambda'^{+}_{1}(q'^\mu_{_P\perp})=
\Lambda'^{-}_{1}(q'^\mu_{_P\perp})\frac{(\not\!{
q'}_{P\perp}+\alpha_1\not\!{P'}_{P\perp})}{\omega'_1} \,,\;\;
\Lambda^{-}_{1}(q^\mu_{_P\perp})\Lambda^{+}_{1}(q^\mu_{_P\perp})=
\Lambda^{-}_{1}(q^\mu_{_P\perp})\frac{\not\!{q}_{P\perp}}{\omega_1}\,.
\end{equation}

When the photon is emitted by the $2$-quark  i.e. the $1$-quark is
kept unchanged (Fig.2), similar relations may be obtained. Namely
the photon is emitted by the anti-quark, i.e. there is the
condition
$\delta^3({q}_{P\perp}-\alpha_1{P'}_{P\perp}-{q'}_{P\perp})$, then
we have $\omega'_1=\omega_1$,
$\Lambda'^{\pm}_{1}=\Lambda^{\pm}_{1}$ and the
$\Lambda'^{\pm}_{2P}(q_{P_\perp})$, then the identities:
\begin{equation}\label{appa14}
\Lambda'^{\pm}_{2}(q'^\mu_{_P\perp})\Lambda'^{\pm}_{2}(q'^\mu_{_P\perp})=
\mp\frac{{m_2}}{\omega'_2}\Lambda'^{\pm}_{2}(q'^\mu_{_P\perp})
\,,\;\;
\Lambda^{\pm}_{2}(q^\mu_{_P\perp})\Lambda^{\pm}_{2}(q^\mu_{_P\perp})=
\mp\frac{{m_2}}{\omega_2}\Lambda^{\pm}_{2}(q^\mu_{_P\perp})\,;
\end{equation}
\begin{equation}\label{appa15}
\Lambda'^{+}_{2}(q'^\mu_{_P\perp})\Lambda'^{-}_{2}(q'^\mu_{_P\perp})=
-\frac{(\not\!{q'}_{P\perp}-\alpha_2\not\!{P'}_{P\perp})}
{\omega'_2}\Lambda'^{-}_{2}(q'^\mu_{_P\perp}) \,,\;\;
\Lambda^{+}_{2}(q^\mu_{_P\perp})\Lambda^{-}_{2}(q^\mu_{_P\perp})=
-\frac{\not\!{q}_{P\perp}}{\omega_2}\Lambda^{-}_{2}(q^\mu_{_P\perp})\,;
\end{equation}
\begin{equation}\label{appa16}
\Lambda'^{-}_{2}(q'^\mu_{_P\perp})\Lambda'^{+}_{2}(q'^\mu_{_P\perp})=
-\Lambda'^{-}_{2}(q'^\mu_{_P\perp})\frac{(\not\!{q'}_{P\perp}
-\alpha_2\not\!{P'}_{P\perp})}{\omega'_2} \,,\;\;
\Lambda^{-}_{2}(q^\mu_{_P\perp})\Lambda^{+}_{2}(q^\mu_{_P\perp})=
-\Lambda^{-}_{2}(q^\mu_{_P\perp})\frac{\not\!{q}_{P\perp}}{\omega_2}\,.
\end{equation}

\end{document}